\documentclass[traditabstract]{aa} 
\usepackage{graphicx}
\usepackage{hyperref}
\usepackage{latexsym}
\usepackage{color}
\usepackage{amsmath,amssymb}

\usepackage{natbib}
\usepackage{soul}
\usepackage{ulem}
\usepackage{txfonts}
\usepackage{float}
\usepackage{longtable,lscape}
\newcommand{\degree}{$^{\circ}$}
\newcommand{\teff}{$T_{\rm eff}$}
\newcommand{\logg}{$\log g$}
\newcommand{\vsini}{$v\sin i$}

\newcommand{\rotfit}{{\sf ROTFIT}}

\newcommand{\COMPO}{{\sf COMPO2}}
\newcommand{\kms}{km\,s$^{-1}$}
 
\newcommand{\gaia}{{\it Gaia}}
\newcommand{\tess}{{\it TESS}}

\newcommand{\jktebop}{{\sf JKTEBOP}}
\newcommand{\eagles}{{\sf EAGLES}}

\newcommand{\Msun}{{$M_{\odot}$}}
\newcommand{\Lsun}{{L$_{\odot}$}}
\newcommand{\Rsun}{{$R_{\odot}$}}

\newcommand{\Mstar}{{$M_{\ast}$}}
\newcommand{\Lstar}{{$L_{\ast}$}}
\newcommand{\Rstar}{{$R_{\ast}$}}

\newcommand{\tic}{{TIC~43152097}}

\definecolor{blu}{rgb}{0,0,1}
\definecolor{mag}{rgb}{1,0,1}

\begin{document}

\title{TIC~43152097\\  The first eclipsing binary in NGC~2232 \thanks{Based on observations made with the Italian Telescopio Nazionale Galileo (TNG), 
operated on the island of La Palma by the Fundaci\'on Galileo Galilei of the INAF (Istituto Nazionale di Astrofisica) at the Observatorio del Roque 
de los Muchachos. 
Based on photometry collected at the INAF - Osservatorio Astrofisico di Catania.}}  

\author{A. Frasca\inst{\ref{instOACT}} 
\and    J. Alonso-Santiago\inst{\ref{instOACT}}
\and    G. Catanzaro \inst{\ref{instOACT}}
\and    A. Bragaglia\inst{\ref{instOABO}} 
\and    V. D'Orazi\inst{\ref{instUnitov},\ref{instOAPD}}
\and    X. Fu\inst{\ref{instPurpleMount},\ref{instOABO}}
\and    A. Vallenari\inst{\ref{instOAPD}}
\and    G. Andreuzzi\inst{\ref{instTNG},\ref{instOARM}}
} 

\institute{INAF -- Osservatorio Astrofisico di Catania, via S. Sofia 78, 95123 Catania, Italy \label{instOACT}\\ \email{antonio.frasca@inaf.it}\label{OACT}
\and INAF -- Osservatorio di Astrofisica e Scienza dello Spazio, Via P. Gobetti 93/3, 40129 Bologna, Italy \label{instOABO}
\and
Department of Physics, University of Rome Tor Vergata, via della Ricerca Scientifica 1, 00133, Rome, Italy\label{instUnitov}
\and
INAF -- Osservatorio Astronomico di Padova, Vicolo dell\u2019Osservatorio 5, 35122, Padova, Italy\label{instOAPD}
\and
Purple Mountain Observatory, Chinese Academy of Sciences, Nanjing 210023, China\label{instPurpleMount}
\and
Fundaci\'on Galileo Galilei--INAF, Rambla Jos\'e Ana Fern\'andez P\'erez 7, 38712 Bre\~na Baja, Tenerife, Spain\label{instTNG}
\and
INAF -- Osservatorio Astronomico di Roma, Via Frascati 33, 00078 Monte Porzio Catone, Italy\label{instOARM}
}

\date{Received 19 June 2023 / Accepted 24 July 2023}

\abstract{We report the discovery of a low-mass totally eclipsing system in the young ($age\simeq28$\,Myr) open cluster NGC\,2232, during an examination of their \tess\ photometry. The follow-up study of this detached system, \tic, is based on photometry and high-resolution spectra from the literature and collected by us. The radial velocity of the center of mass and the photospheric lithium abundance of the binary components confirm its membership to NGC\,2232. By analyzing the existing photometric and spectroscopic data, we obtain orbital elements and fundamental stellar parameters for the two stars. The primary component of \tic\ is a late F-type dwarf (\teff\,=\,6070\,K), while the lower-mass secondary is a late K-type star (\teff\,=\,4130\,K) that is still in the pre-main-sequence phase. The precise measurements of the radii, masses, and effective temperatures, enabled by the simultaneous solution of light and radial velocity curves, indicate radius inflation for the K-type component, which turns out to be 7--11\,\% larger than that predicted by standard evolutionary models. More sophisticated models incorporating both the inhibition of convective energy transport caused by sub-photospheric magnetic fields and the effects of cool starspots covering a substantial fraction of the stellar surface (30--60\,\%) allow the position of the secondary component to be reproduced in the Hertzsprung-Russell and mass-radius diagrams.  
} 

   \titlerunning{Parameters of the first eclisping binary in NGC~2232}
      \authorrunning{A. Frasca et al.}

\maketitle

\section{Introduction}
\label{Sec:intro}

Precise measurements of stellar parameters are of paramount importance for studying the physics and evolution of stars and their environment, specifically, circumstellar protoplanetary disks and planetary systems.
Eclipsing binaries represent unparalleled targets for achieving the most precise determinations of fundamental stellar parameters, such as effective temperature (\teff), mass (\Mstar), and radius (\Rstar). These quantities can be directly measured from the analysis of the radial velocity (RV) and light curves, eliminating the need for calibration relationships or  complex models of stellar structure and evolution. Furthermore, these precisely derived parameters serve as a means to validate the accuracy and reliability of internal structure models, providing a valuable opportunity to assess the fidelity of our understanding of stellar interiors.

One of the most relevant results that emerged from the study of binaries with late spectral type (KM) components  is the disagreement between the theoretical and observed radii of young magnetically active stars.
For a fixed mass, the observed radii are 10--20\% larger than the model predictions and hence, for a given luminosity, the effective temperature can be overestimated by $\sim$5\% \citep[e.g.,][]{Morales2009,Kraus2011,Torres2013}. 
On the other hand, the radii of single slowly rotating stars derived from interferometric measurements \citep[e.g.,][]{Demory2009,Boyajian2012} are in close agreement with standard models, suggesting that the high level of magnetic activity induced by tidal coupling is the main effect responsible for the radius inflation in the components of close binary systems. Strong support for this hypothesis comes from observations of late-type members of young clusters, which have not yet spun down via magnetic braking and display a strong magnetic activity. 
Their color-magnitude diagrams (CMD) and lithium abundance distributions are reproduced satisfactorily, considering only radius inflation of approximately 10\% \citep[e.g.,][]{Jeffries2017,SomersStassun2017,Jackson2018}.

Eclipsing binaries with components in the pre-main sequence (PMS) or in the zero-age main sequence (ZAMS) phase, and in particular those belonging to clusters with a known age and metallicity,  are of great help in this respect, because they offer the possibility for direct measures of \teff, \Mstar, and \Lstar\ for objects in these evolutionary phases. This is particularly relevant for PMS binaries, because the components of the  already known binaries with radius inflation are mostly main-sequence (MS) stars.

\begin{figure}
\begin{center}
\includegraphics[width=9.5cm]{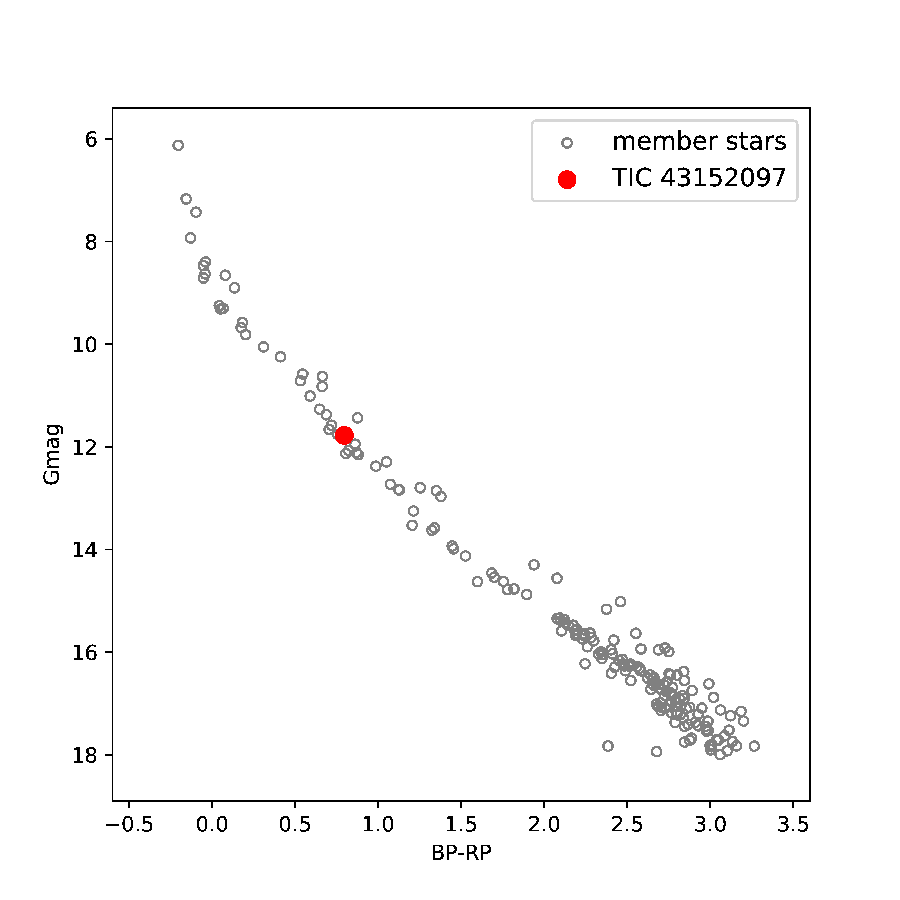}     
\caption{\gaia\ CMD for the members of NGC\,2232 (open circles) selected by \citet{Cantat2020}. The position of \tic\ is marked with a red dot.}
\label{Fig:GaiaCMD}
\end{center}
\end{figure}

We report here the discovery and study of an eclipsing binary, \tic, in NGC~2232, a young open cluster ($\text{age}\simeq$\,28--38\,Myr) with a solar metallicity \citep[e.g.,][]{Binks2021,Jeffries2023}, whose late-type members are still in the PMS phase.
\tic\ was initially selected as a target for high-resolution spectroscopic observations of members of young open clusters tentatively associated with the  Radcliffe Wave \citep{Alves2020}, with the aim of determining their kinematic and chemical properties. 
For this purpose, we selected slowly rotating G-type members of five clusters, including NGC~2232, from the high-probability candidates proposed by \citet{Cantat2018,Cantat2020}.
Therefore, among the stars with Transiting Exoplanet Survey Satellite (TESS; \citealt{Ricker15}) photometry, we searched for those with rotation periods long enough to expect a small projected rotation velocity (\vsini\,$\leq20$\,\kms), which would  allow  a good determination of atmospheric parameters and abundances to be made.
In the course of this search, we came across an eclipsing binary that was not known in the literature and that  had all the characteristics of a bona fide member of NGC~2232, as clearly shown by its position in the \gaia\ CMD (Fig.\,\ref{Fig:GaiaCMD}). The main properties of this source from the literature are summarized in Table~\ref{Tab:properties}.
We acquired three spectra of this star during our observing run with High Accuracy Radial velocity Planet Searcher North spectrograph (HARPS-N), which, together with another spectrum available in the literature, are well enough distributed in the orbital phase to trace the RV curve. 
We solved for the \tess\ light curve and the RV curve to obtain the orbital and physical parameters of this system.

The paper is organized as follows. In Sect.~\ref{Sec:Observations} ,we present our observations and the data retrieved from the literature.
In Sect.~\ref{Sec:Results}, we show the results of our work, describing the analysis carried out on both the photometric and spectroscopic data. 
We discuss our results in Sect.~\ref{Sec:discussion}. 
Finally, in Sect.~\ref{sect:Conclusions}, we summarize the main results and present our conclusions.

\section{Observations}  
\label{Sec:Observations}

\begin{table*}
\caption{Properties of \tic\ from the literature.}
\begin{center}
\begin{tabular}{ccccccccccc}
\hline
\hline
\noalign{\smallskip}
\gaia\ DR3 ID &  RA         &  DEC   & Plx$^a$ & pmRA & pmDEC   & $G^a$   & $G_{\rm BP}-G_{\rm RP}^a$ & $V^b$ & $B-V^b$  \\
              & (J2000)     &(J2000) & (mas) & \multicolumn{2}{c}{(mas\,yr$^{-1}$)}  & (mag) & (mag) & (mag) & (mag)     \\
\noalign{\smallskip}
\hline
\noalign{\smallskip}
3104591653648488192 & 06 29 38.05 & $-$04 03 59.2 &  2.9119 & $-4.556$  &       $-1.864$   & 11.771  & 0.791  &  11.871 & 0.613  \\  
\noalign{\smallskip}
\hline
\end{tabular}
\end{center}
{\bf Notes.} $^a$ From the \gaia~DR3 catalog \citep{GaiaDR3}. $^b$ From the {\it APASS} Catalog \citep{Henden2018}.
\label{Tab:properties}
\end{table*}

\begin{figure}
\begin{center}
\includegraphics[width=9.4cm]{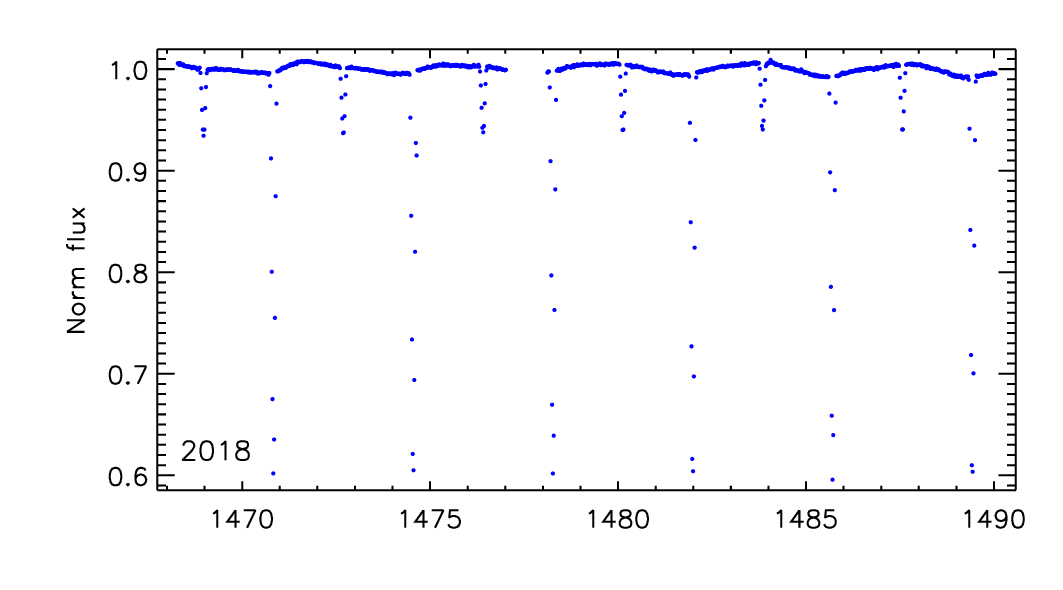}    
\includegraphics[width=9.4cm]{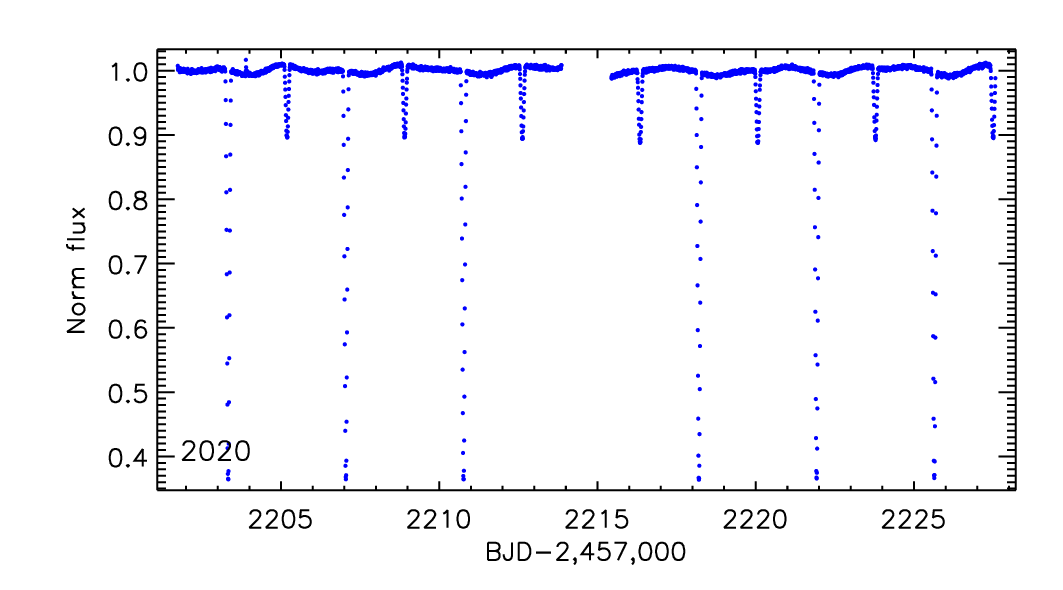}    
\caption{SAP {\it TESS} light curves of TIC~43152097 in 2018 ({\it top}) and 2020 ({\it bottom}). Low-amplitude out-of-eclipse variation is visible in both epochs. The higher data cadence in 2020 is apparent.}
\label{fig:TESS}
\end{center}
\end{figure}

\subsection{Photometry}
\label{Subsec:Obs_photo}

Space-born accurate photometry was obtained with \tess\ \citep{Ricker15}.
TIC~43152097 was observed in sector 6 with 1800\,s exposure times between 15\ December 2018 and 6\ January 2019,
as well as in sector 33 with 600\,s exposure times between 17\ December 2020 and 12 January 2021.
As NGC\,2232 is a sparse cluster, there is no severe star crowding, particularly  considering the large pixel size of \tess\ (21\arcsec). Moreover, TIC\,43152097 is located far from the central region of the cluster, at about 40\,\arcmin\ from 10\,Mon, which is the brightest star of the cluster.
Indeed, searching in the \gaia~DR3 catalog \citep{GaiaDR3}, no star with a comparable magnitude can be found within 21\arcsec of TIC\,43152097.
We downloaded the \tess\ light curves reduced by the MIT Quick Look Pipeline \citep[QLP,][]{Huang2020} from the MAST\footnote{\url{https://mast.stsci.edu/portal/Mashup/Clients/Mast/Portal.html}} archive and used the simple aperture photometry (SAP) flux and the KSPSAP flux. The latter was obtained from the SAP flux by applying  a high-pass filter that removes the low-frequency variability originating from stellar activity or instrumental effects.

To obtain the color information that is lacking in the \tess\ data, we planned multiband photometric observations of TIC~43152097 at the {\it M. G. Fracastoro} station (Serra La Nave, Mt. Etna,
1750 m. a.s.l.) of the {\it Osservatorio Astrofisico di Catania} (OACT, Italy) in the winter season of 2022-23. 
We used the facility imaging camera at the 0.91\,m telescope with a set of broadband Bessell filters ($B$, $V$, $R_{\rm C}$, and  $I_{\rm C}$). 
Due to bad weather and time constraints, we were only able, apart from some scattered data points outside eclipses, to acquire useful data during a secondary eclipse on 7 January 2023 whose egress, observed at high values of airmass, was not fully covered. Although their cadence is higher than that of \tess, the photometric precision is lower. Therefore, we used this multiband photometry only to verify whether the light-curve solution made on the \tess\ data was able to reproduce them (see Fig.\,\ref{fig:ground_LC}).

\subsection{Spectroscopy}
\label{Subsec:Obs_spe}

High-resolution  spectroscopy ($R\simeq115,000$) was performed  with the High Accuracy Radial velocity Planet Searcher North spectrograph
(HARPS-N, \citealt{Cosentino2012}), which is mounted at the 3.6-m \textit{Telescopio Nazionale Galileo} (TNG), at the Roque de los Muchachos Observatory (La Palma, Spain). 
Three spectra were acquired during three consecutive nights from 27 to 29 October 2022, with 1800\,s exposure time. These spectra were automatically reduced using the instrument pipeline.

Searching in the literature for additional high-resolution spectra, we found only one spectrum, which is the sum of three individual spectra taken on the same night with HERMES at AAT in the framework of the Galactic Archeology with HERMES survey (GALAH), \cite{desilva2015}, \citep[see][for the third GALAH release]{Buder2021}. HERMES \citep{sheinis2016} is a multi-object spectrograph ($R\simeq28,000$) mounted at the 3.9-m telescope of the Anglo Australian Observatory. The secondary component was not detected by the automatic pipeline, which has basically derived the atmospheric parameters, element abundances, and RV of the primary component of \tic. 
We used this spectrum to measure the RV of the system components, adding another useful point to their RV curves. 

The \gaia\ DR3 catalog \citep{GaiaDR3} reports an RV=36.46\,$\pm$\,19.43\,\kms, which was obtained with the Radial Velocity Spectrometer (RVS) as the mean of the values measured in eight different epochs. The large error for such a bright source is suggestive of a variable RV. Unfortunately, the individual values of RV, which would have been helpful to cover the RV curve of the primary component, will only be available in the fourth \gaia\ data release.

\section{Results}
\label{Sec:Results}

\subsection{Light-curve solution}  
\label{Subsec:photo}

The SAP \tess\ light curves (Fig.~\ref{fig:TESS}) clearly show that this star is indeed an eclipsing binary with the primary minimum much deeper than the secondary one, indicating a very different temperature for the two system components. 
In the SAP data, a low-amplitude variation is also visible outside the eclipses that can be ascribed to starspots in one of the components and to instrumental effects.
For the purpose of studying the eclipses, it is advantageous to use the KSPSAP flux, in which low-frequency variations, whatever their origin, have been filtered out and the out-of-eclipse part of the light curve appears flat.

The first parameter we measured was the orbital period, $P_{\rm orb}$.
To this aim, we determined the times of primary and secondary minima with the method of eclipse bisector \citep[e.g.,][]{Covino2004}, which allowed us to reduce the uncertainties on the mid-eclipse epochs due to the measurement errors and the coarse data sampling (see Fig.~\ref{Fig:bisector} for an example). 
The errors of the mid-eclipse epochs, measured as the standard deviation of the values of eclipse bisector, range from approximately 25 to 300\,s, depending on 
the signal-to-noise ratio (S/N) and the data cadence.  

The epochs of the primary and secondary minima allowed us to find the constant-period ephemeris, which provided the best match to the data (minimum residuals):
\begin{equation}
    BJD_{\rm minI} = 2\,459\,225.6410(1) + 3.718265(7) \times E. 
    \label{Eq:ephem}
\end{equation}

\begin{figure}
\begin{center}
\hspace{-.5cm}
\includegraphics[width=9.cm]{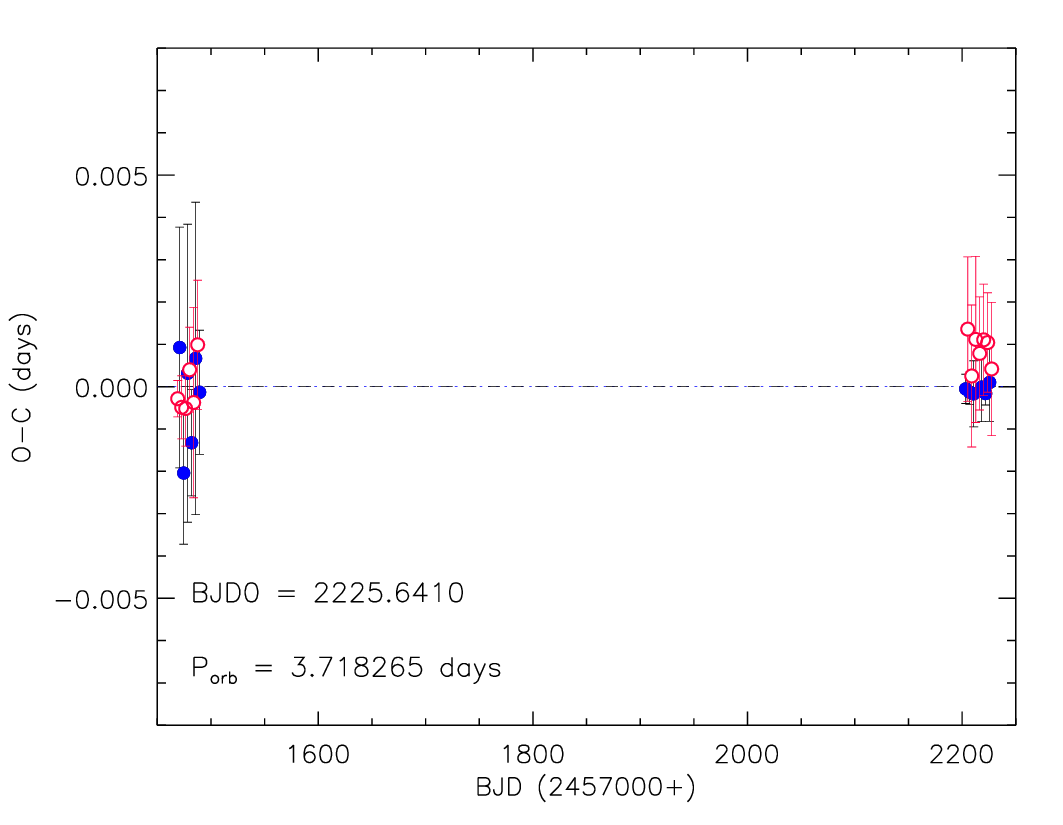} 
\vspace{0.cm}
\caption{Observed minus calculated times of minima for TIC~43152097. Primary and secondary minima are represented with blue dots and red circles, respectively. The orbital period appears to be constant within the errors. 
}
\label{Fig:o-c}
\end{center}
\end{figure}

{\noindent The residuals between observed and calculated epochs (O--C) of primary (blue dots) and secondary (red circles) minima (Fig.~\ref{Fig:o-c}) are flat and scattered
around zero, which means no relevant period variation occurred during the time spanned by the \tess\ observations.
A very small systematic offset between the primary and secondary minima observed in sector 33 seems to be visible in the O--C, which could suggest a very small eccentricity or could be the effect of an uneven distribution of starspots in the secondary component. However, this is well within the error bars, and it is likely to be not significant.}

\begin{figure*}[t]
\includegraphics[width=9.cm]{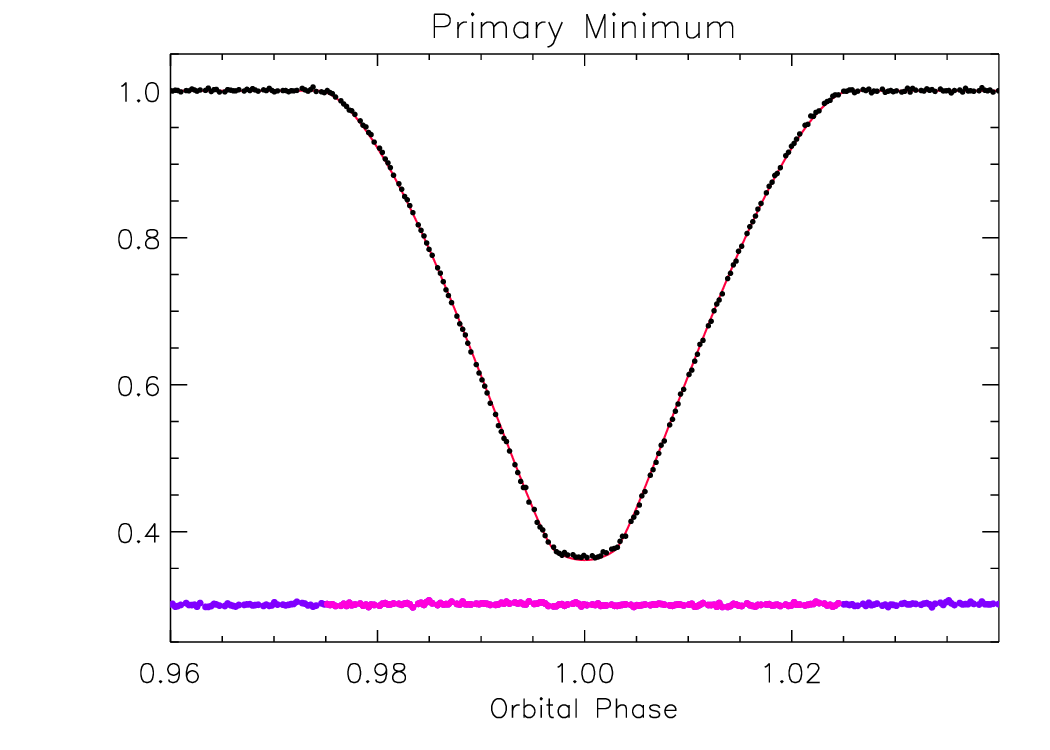}     
\includegraphics[width=9.cm]{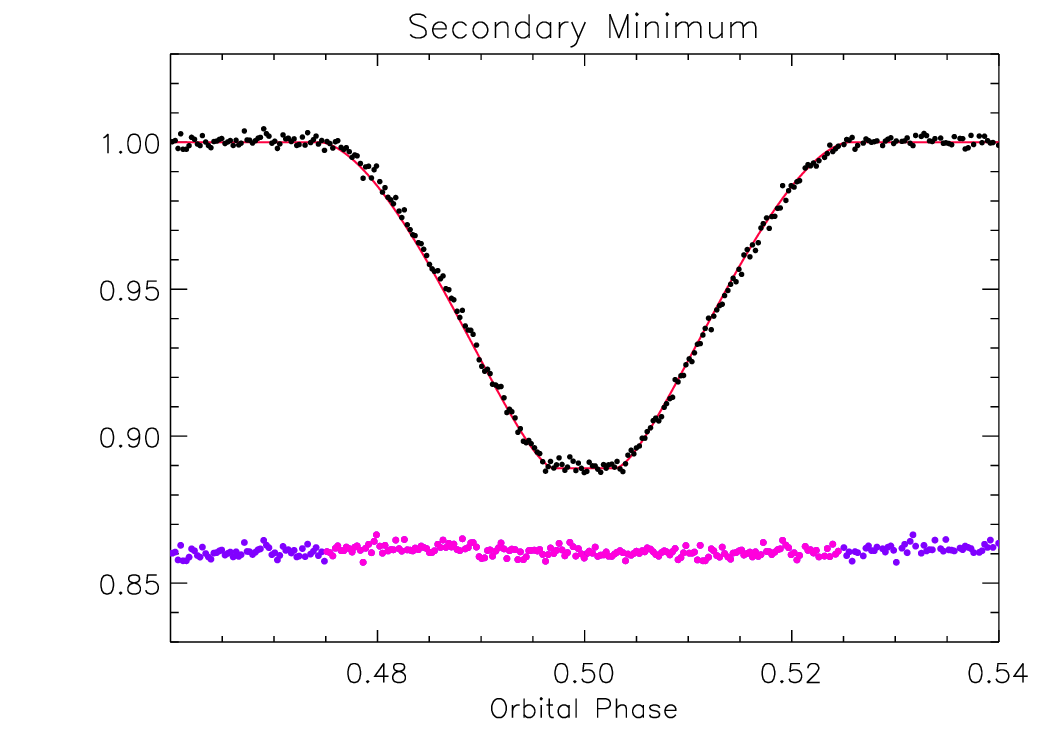}     
\caption{Primary ({\it left panel}) and secondary ({\it right panel}) eclipse in the phased KSPSAP {\it TESS} light curve of sector 33 (black dots). The solid red lines represent the solution, and the blue dots at nearly zero level are the residuals (observed$-$model). The magenta dots are those used to calculate the $\chi^2$ of the fit.}
\label{fig:eclipses}
\end{figure*}

We used the KSPSAP \tess\ data from 2020, which have a higher cadence, for the light-curve solution. The phase-folded light curve around the primary and secondary minimum is shown in Fig.~\ref{fig:eclipses} (black dots), along with the synthetic light curve (solid red lines).
Given the large separation of the two components, suggested by the short duration of the eclipses in units of the orbital period, we adopted a model with spherical limb-darkened stars.
We used the quadratic limb-darkening coefficients calculated for the \tess\ passband  by \citet{Claret2017}, who adopted quasi-spherical PHOENIX-COND stellar atmosphere models with solar metallicity, \logg\,=\,4.5, and mixing-length parameter $\alpha=2$. 
The occulted fraction of the limb-darkened star disk at each orbital phase was calculated from numerical integrations following the prescriptions of \citet{Kopal1990} and dividing the stellar disk into 1000 concentric rings. 
The surface brightness ratio between the two stars, $J$, was calculated, as a function of \teff, by integrating BT-Settl models \citep{Allard2012} through the \tess\ passband.
We fixed the temperature of the primary component as $T_1=6070\pm 70$\,K (see Sect.\,\ref{Subsec:param}) and left the temperature $T_2$ of the secondary component free to vary. The system inclination, $i$, and the fractional radii of the primary and secondary components in units of the separation, $r_1=R_1/a$ and $r_2=R_2/a$,
were also free variables.
We minimized the residuals in the part of the light curve encompassing the two eclipses (magenta dots in Fig.~\ref{fig:eclipses}), using as a goodness-of-fit parameter the $\chi^2$ defined as
\begin{equation}
    \chi^2 = \frac{1}{N} \sum \left(\frac{F_{\rm obs}-F_{\rm mod}}{\sigma_{\rm obs}} \right)^2 ,
\end{equation}

\noindent{where $N$ is the total number of points selected for the fit, $F_{\rm obs}$ and $F_{\rm mod}$ are the observed and model fluxes,  
respectively, and $\sigma_{\rm obs}$ is the data uncertainty.}

\begin{table}   
\begin{center}
\caption{Best-fit parameters from the model solution of the KSPSAP  \tess\ light curve of \tic\ in 2020.}
\begin{tabular}{lr}
\hline
\hline
\noalign{\smallskip}
Parameter &  Best-fit value \\ 
(units)   &      \\
\hline
\noalign{\smallskip}
BJD$_0^{(a)}$        &  2225.641$\pm$0.001     \\
$P_{\rm orb}$ ($d$) &  3.718265$\pm$0.000007  \\
$e$                 &  0                      \\
$\omega$ (\degree)  &  \dots                  \\
$i$      (\degree)  &  $90^{+0}_{-0.4}$       \\
$r_1$               &  0.0894$\pm$0.0006      \\
$r_2$               &  0.0706$\pm$0.0004      \\
$T_1^{(b)}$ (K)     &  6070$\pm$70            \\
$T_2$ (K)           &  4130$\pm$30            \\
$a_1$, $b_1^{(c)}$  & 0.3276, 0.2297          \\
$a_2$, $b_2$        & 0.3955, 0.2618          \\
$J^{(d)}$           & 0.194                   \\
\noalign{\smallskip}
\hline \\
\end{tabular}
\label{Tab:LC_Param}
~\\
{\bf Notes.} $^{(a)}$ Barycentric Julian date (BJD-2\,457\,000) of the primary minimum; the subscripts 1 and 2 refer to the primary (more massive) and secondary component, respectively. $^{(b)}$ $T_1$ fixed based on spectral analysis. 
$^{(c)}$ Quadratic limb-darkening coefficients. $^{(c)}$ Surface brightness ratio $J=F_2/F_1$ calculated in the \tess\ passband from BT-Settl models.
\end{center}
\end{table}

We also tried a solution with linear limb darkening for both stars and found a slightly worse fit, especially at eclipse contacts.
However, the parameters of the final model changed very little:$i$ and $T_2$ did not change, $r_1=0.0890$ (approximately 0.4\% smaller), and $r_2=0.0725$ (2.7\% larger). 
The contour maps of the $\chi^2$ for various pairs of parameters are shown in Fig.\,\ref{fig:chi2}, where the 1$\sigma$ confidence level is marked with a thick red line. 
The best-fit parameters are listed in Table\,\ref{Tab:LC_Param}, together
with their standard errors, which correspond to the 1$\sigma$ confidence
levels.

We used also the v43 version of the {\sc fortran} code \jktebop\footnote{\url{https://www.astro.keele.ac.uk/jkt/codes/jktebop.html}} 
\citep{Southworth2004MNRAS,Southworth2013} to verify our results. 
This code uses a biaxial ellipsoidal model \citep{NelsonDavis1972} and is particularly suitable for very fast modeling of detached
eclipsing binaries. We first ran \jktebop\ with the same assumptions of our code, that is, considering spherical stars with a quadratic limb-darkening law with \citet{Claret2017} coefficients. The stellar radii, the system inclination, and 
the surface brightness ratio were left as free parameters in the fit; the results are very similar to those of our code,  $r_1=0.0898\pm 0.0025$, $r_2=0.0730\pm 0.0024$, $i=90.0^{+0.0}_{-0.8}$\,degrees, and $J=F_2/F_1=0.190\pm 0.003$. The 1$\sigma$ uncertainties in the fitted parameters were determined 
by means of 1000 Monte Carlo simulations.
The radius of the secondary component
is only 3.4\% larger than the one derived with our code. Relaxing the hypothesis of spherical stars and leaving $P_{\rm orb}$ and BJD$_0$ free to vary did not  
appreciably change the final parameters. We also tried with an eccentric orbit, but this did not improve the fit. Another trial was performed leaving the third light, $L_3$, as a 
free parameter and we in fact found a better fit, with a small negative value, $L_3=-0.094\pm 0.007$. Small $L_3$ values have been found  in the solution of \tess\ light 
curves \citep[e.g.,][]{Southworth2022} and could be related to sky background estimation. If, on the other hand, this third light were to be real, it would leave  $r_1$ nearly 
unchanged and would have a detectable effect only on the flux ratio ($J=0.1951\pm 0.0007$) and on the radius of the secondary component ($r_2=0.06895\pm 0.0001$), 
making it closer to and slightly smaller ($\sim$\,2\%) than the value reported in Table~\ref{Tab:LC_Param}.
In the following section, we use the parameters derived with our model and reported in Table~\ref{Tab:LC_Param}, but we also consider the values of $r_2$ derived with 
\jktebop,\ both assuming $L_3=0$ and adopting the best-fit value.

\subsection{Radial velocity}
\label{Subsec:RV}
The RV was measured by cross-correlating the target spectra with late-type stellar templates, which are {\sc ATLAS9}  \citep{Kurucz1993}
synthetic spectra with solar metallicity and \teff\ in the range 4000--6000\,K calculated with {\sc SYNTHE} \citep{Kurucz1981}.  
This analysis was carried out with the IRAF\footnote{IRAF is distributed by the National Optical Astronomy Observatory, which is operated by the Association of 
Universities for Research in Astronomy, Inc.} task {\sc fxcor}, excluding very broad features that can blur the peaks of the cross-correlation 
function (CCF). 
To measure the centroids of the CCF peaks of the two components, we applied a two-Gaussian fit. 
The RV error, $\sigma_{\rm RV}$, was computed by {\sc fxcor} according to the fitted peak height and the antisymmetric noise as described by \citet{Tonry1979}.

\setlength{\tabcolsep}{4pt}

\begin{table}   
\caption{Barycentric RVs of the two components of TIC~43152097.}
\begin{tabular}{lrcrclr}
\hline
\hline
\noalign{\smallskip}
 BJD     &  RV$_{\rm 1}$   &  $\sigma_{\rm RV_1}$ &  RV$_{\rm 2}$   &  $\sigma_{\rm RV_2}$ & Instrument & S/N$^{(a)}$ \\
(2\,450\,000+) &  \multicolumn{2}{c}{(\kms)} & \multicolumn{2}{c}{(\kms)} & \\
 \hline
\noalign{\smallskip}
 8145.0608 & $-17.15$ & 0.43 &   93.70  & 2.83 & HERMES  & 105 \\
 9879.7243 &    61.54 & 1.64 & $-30.26$ & 4.06 & HARPS-N &  41 \\
 9880.7015 & $-33.89$ & 0.27 &  118.17  & 2.73 & HARPS-N &  43 \\ 
 9881.6640 &  $-1.66$ & 0.38 &   68.52  & 1.08 & HARPS-N &  33 \\
\noalign{\smallskip}
\hline \\
\end{tabular}
\label{Tab:RV}
~\\
{\bf Notes.} $^{(a)}$ Signal-to-noise ratio per pixel at 6500 \AA.
\end{table}

\begin{figure}
\begin{center}
\hspace{-.5cm}
\includegraphics[width=9.cm]{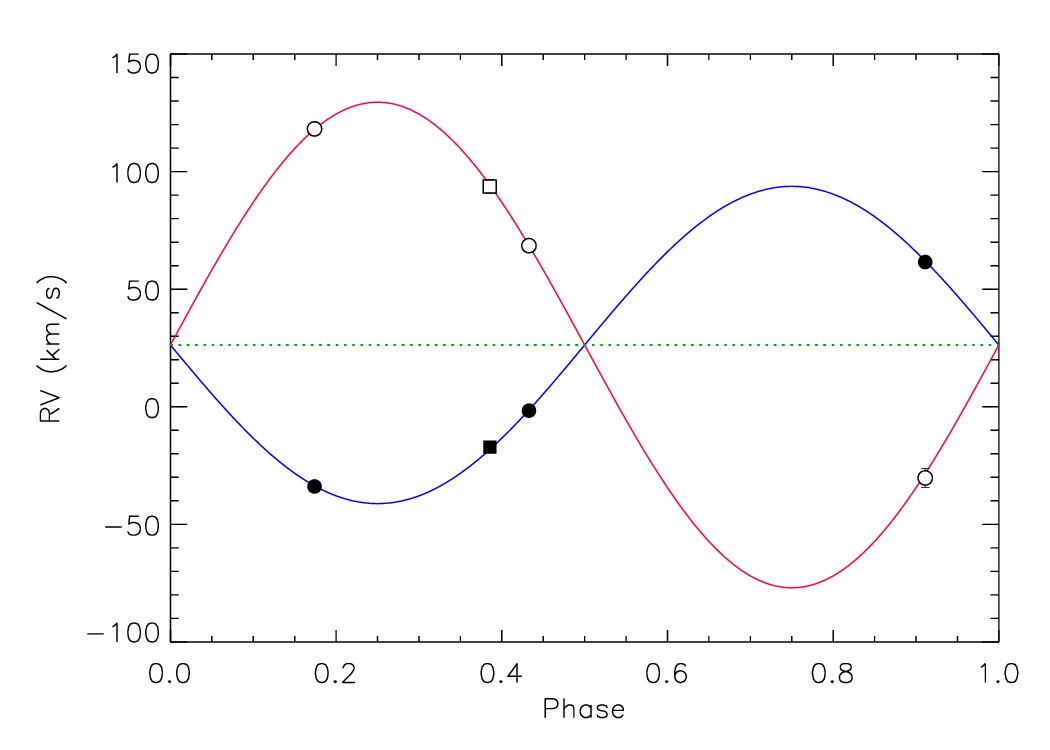}      
\vspace{0.cm}
\caption{Barycentric RV curve (circles = HARPS-N, squares = HERMES)
of \tic. Filled and open symbols are used for the primary (more massive) and secondary component, respectively. The blue and red lines represent the orbital solution (Table\,\ref{Tab:Param}) 
for the primary and secondary component, respectively. 
}
\label{Fig:RV_curve}
\end{center}
\end{figure}

The individual values of RV measured for the two system components are listed in Table~\ref{Tab:RV}. These data, folded with the ephemeris in Eq.~\ref{Eq:ephem}, are displayed in  Fig.~\ref{Fig:RV_curve} along with the circular orbital solution. The latter was fitted to the data by means of the {\sc curvefit} routine \citep{Bevington}, allowing us to determine the orbital parameters and their standard errors, which are reported in Table~\ref{Tab:Param}.

\begin{table}   
\begin{center}
\caption{Orbital and stellar parameters of TIC~43152097.}
\begin{tabular}{lr}
\hline
\hline
\noalign{\smallskip}
\multicolumn{2}{c}{Orbital parameters}   \\ 
\hline
\noalign{\smallskip}
 BJD$_0^{(a)}$        &  2225.641$\pm$0.001         \\
 $P_{\rm orb}$ ($d$) &  3.718265$\pm$0.000007      \\
 $e$                 &  0                          \\
 $\omega$ (\degree)  &  \dots                      \\
 $\gamma$ (\kms)     &  26.27$\pm$0.58 \\ 
 $K_{\rm 1}$ (\kms)  &  67.48$\pm$0.78 \\ 
 $K_{\rm 2}$ (\kms)  & 103.28$\pm$3.32 \\ 
 $M_{\rm 1}\sin^3i$ (M$_{\sun}$) & 1.160$\pm$0.083 \\ 
 $M_{\rm 2}\sin^3i$ (M$_{\sun}$) & 0.758$\pm$0.033 \\ 
 $q=M_{\rm 2}/M_{\rm 1}$  &  0.653$\pm$0.022 \\ 
 $a\sin i$ (R$_{\sun}$) & 12.54$\pm$0.25  \\ 
\noalign{\smallskip}
\hline
\noalign{\smallskip}
\multicolumn{2}{c}{Stellar parameters}   \\ 
\hline
\noalign{\smallskip}
 M$_{\rm 1}$ (M$_{\sun}$) & 1.160$\pm$0.083 \\ 
 M$_{\rm 2}$ (M$_{\sun}$) & 0.758$\pm$0.033 \\ 
 R$_{\rm 1}$ (R$_{\sun}$) & 1.121$\pm$0.022 \\ 
 R$_{\rm 2}$ (R$_{\sun}$) & 0.885$\pm$0.021 \\ 
\logg$_1$ (cgs)     & 4.423$\pm$0.020   \\
\logg$_2$ (cgs)     & 4.440$\pm$0.016   \\
$T_1$ (K)           &  6070$\pm$70      \\
$T_2$ (K)           &  4130$\pm$60      \\
$L_1$ (\Lsun)       &  1.531$\pm$0.093  \\ 
$L_2$ (\Lsun)       &  0.204$\pm$0.015  \\ 
\hline \\
\label{Tab:Param}
\end{tabular}\\
{\bf Notes.} $^{(a)}$ Barycentric Julian date (BJD-2\,457\,000) of the primary minimum; subscripts 1 and 2 refer to the primary (more massive) and 
secondary component, respectively. 
\end{center}
\end{table}

Combining the light-curve and RV solutions, we derived the absolute parameters of the system components, which are reported in Table~\ref{Tab:Param}. As the inclination is $i=90\degr$, the masses are the same as $M_{1,2}\sin^3i$; the relative errors of the radii are the quadratic sum of the relative errors of the fractional radii, $r_{1,2}$ (reported in Table~\ref{Tab:LC_Param}) and that of the separation, $a$ (Table~\ref{Tab:Param}). The stellar luminosity was calculated as $L=4\pi R^2\sigma$\teff$^4$ and its error was estimated by propagating the \teff\ and $R$ errors. 

\subsection{Stellar parameters from the spectral analysis}
\label{Subsec:param}

As a first approach, we used the code \rotfit\ \citep[]{Frasca2006} to determine the atmospheric parameters and \vsini\ of the brighter component of \tic. For a description of the version of \rotfit\ working on HARPS-N spectra, the reader is referred to \citet{Frasca2019}. We analyzed, for the three spectra, the spectral segments with a wavelength shorter than 5500\,\AA, to minimize the contribution of the cooler secondary component, and found as average parameters of the primary component $T_1=6110\pm80$\,K, \logg$_1=4.18\pm0.13$\,dex, [Fe/H]$_1=0.07\pm0.10$\,dex, and \vsini$=16.0\pm0.9$\,\kms. 

In the case of double-lined spectroscopic binaries (SB2),  we used \COMPO, a code similar to \rotfit\ that was also developed in the {\sf IDL}\footnote{IDL (Interactive Data Language) is a registered trademark of  Harris Corporation.} 
environment \citep{Frasca2006,Frasca2019}.
\COMPO\ adopts a grid of non-active templates to reproduce the observed  spectrum, which is split into segments of 100 \AA\ each that are independently analyzed. 
The grid of templates, as for \rotfit, is composed of ELODIE spectra of low-active slowly rotating FGKM stars.
The resolution of the HARPS-N 
spectra was degraded to that of the ELODIE templates ($R=42,000$) by convolution with a Gaussian kernel with the proper width.
\COMPO\  did not derive the projected rotation velocities of the two components, which are fixed parameters. In this case, with $i=90\degr$, we used the values of \vsini\ that could be derived from the light-curve solution, assuming spin-orbit synchronization (see Sect.\,\ref{Subsec:modulation}), as
\begin{equation}
    v\sin i = v_{\rm eq}\sin i = v_{\rm eq} = \frac{2\pi R_*}{P_{\rm orb}},
\end{equation}
{\noindent finding $v_1\sin i=15.0$\,\kms\ (in  agreement with the values derived with \rotfit) and $v_2\sin i=11.9$\,\kms.}
The RV separation of the two components is well known from the CCF analysis and was used to build the composite "synthetic" spectrum. 
The flux contribution of the primary component in units of the continuum, $w_{\rm 1}$, is normally free to vary in \COMPO. In the case of a totally eclipsing binary such as \tic, we can fix this 
value using the light-curve solution. For instance, at the central wavelength of the $R_{\rm C}$ band, the secondary component contributes to only  7\% of the total flux, that is, $w_{\rm 1}=0.93$; 
at shorter wavelengths, the contribution of the secondary component is still smaller (5\% in the $V$ and 3\% in the $B$ band). Keeping the flux contribution fixed (per each analyzed spectral 
segment), only the templates reproducing the primary and secondary components were variable inside \COMPO.
We ran the code on the spectrum acquired on JD=2\,459\,880.7, which has the largest S/N and RV separation of the components. 
The stellar parameters produced by \COMPO\ for the primary component, namely spectral type F9V, $T_1=6070\pm$70\,K, and \logg$_1$\,=\,4.20$\pm$0.16\,dex,  closely agree with those found with \rotfit. 
Moreover, they share a remarkable level of agreement with the values of \teff=6053$\pm$87\,K and \logg=4.07$\pm$0.22, reported in the GALAH DR3 catalog \citep{Buder2021}.
The atmospheric parameters of the secondary component are only indicative values, given its very low flux contribution and the fairly low S/N of the HARPS-N spectra. 
However, they are compatible with those derived from the light-curve solution, as we found a spectral type K9V, $T_2=4370\pm$450\,K and \logg$_2$\,=\,4.5$\pm$0.3\,dex.

\begin{figure}
\begin{center}
\hspace{-.5cm}
\includegraphics[width=9.0cm]{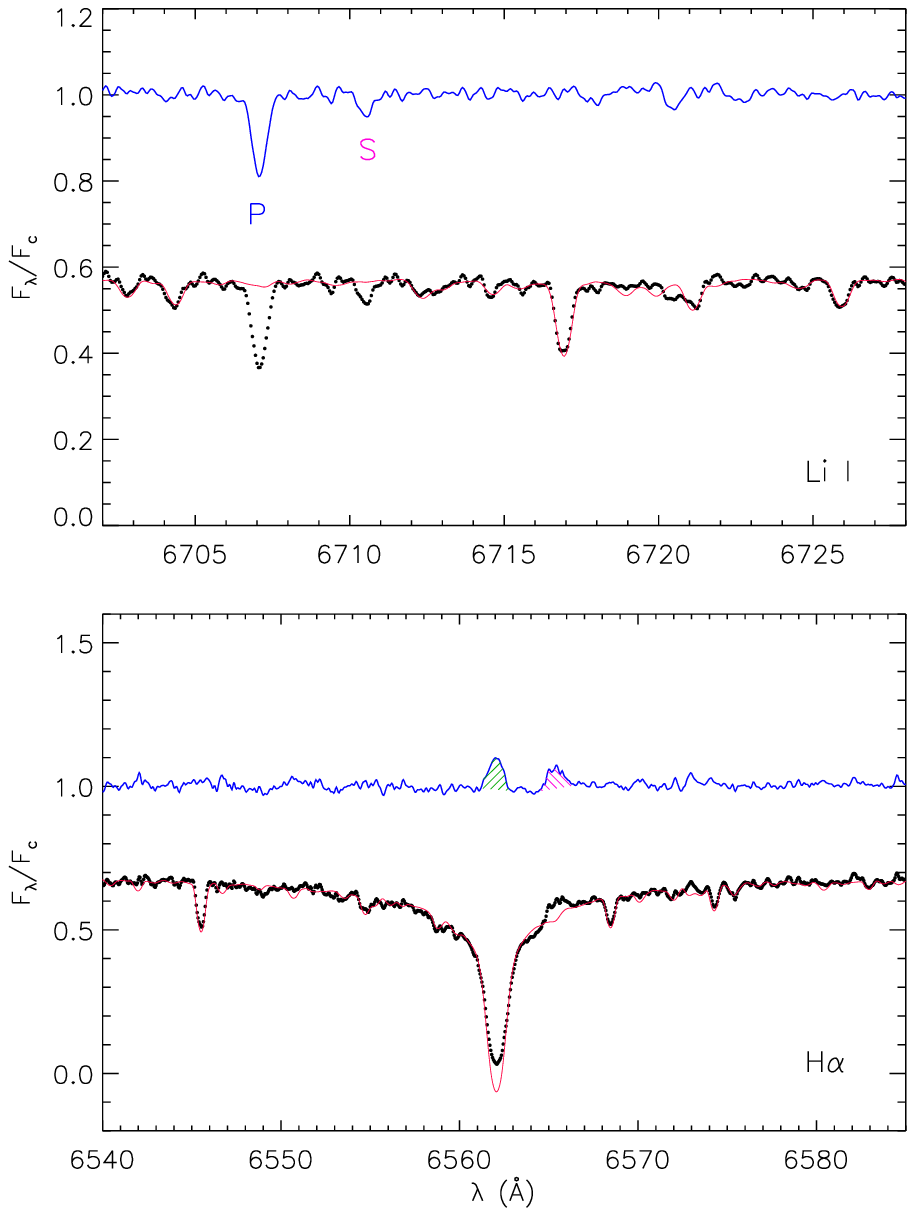}     
\vspace{-1.5cm}
\caption{Subtraction of the synthetic composite spectrum (red line) generated by \COMPO\ from the spectrum of \tic\   observed on 28 October 2022 (black dots). In each panel, the residual spectrum is shown by the blue line.
The chromospheric emission ({\it bottom panel}) in the H$\alpha$ core of the primary and the secondary component, which has been integrated to provide $W_{\rm H\alpha}^{em}$, is shown by the hatched green and magenta  areas, respectively. 
The \ion{Li}{i} $\lambda$6708\,\AA\ absorption lines of the two components ({\it top panel}) are marked with "P" and "S" in the residual spectrum. 
}
\label{fig:subtraction_compo2}
\end{center}
\end{figure}

Equivalent widths of the \ion{Li}{i} 6708\AA\ line, $W_{\rm Li}$, for the  components of TIC~43152097 were measured in the residual spectrum (blue line in Fig.\ref{fig:subtraction_compo2}), which was obtained by subtracting the synthetic template generated by \COMPO\ from the observed spectrum. This procedure offers the advantage of removing possible contamination from nearby iron lines and is particularly helpful in the case of SB2 systems, as it allows us to remove different lines of the two components, which, due to the different Doppler shifts, can overlap the \ion{Li}{i} lines of the two components. 
The error of the equivalent width was estimated as the product of the integration range and the mean error per spectral point, which results from the standard deviation of the flux values of the residual spectrum measured at the two sides of the line.
The $W_{\rm Li}$ values, as well as their errors, were corrected for the flux contribution to the composite spectrum by dividing them by $w_{\rm 1}$ and $w_{\rm 2}$ for the primary and secondary component, respectively. 
Both the measured and corrected values of $W_{\rm Li}$ are listed in Table~\ref{Tab:Halpha_Lithium}.

\begin{table}[htb]
\caption{H$\alpha$, \ion{Li}{i}$\lambda$6708\,\AA\ equivalent widths, and lithium abundance for the components of \tic.}
\begin{center}
\begin{tabular}{lrrrrrlr}
\hline
\hline
\noalign{\smallskip}
Comp    & \teff &  $W_{\rm H\alpha}^{em}$    & err & $W_{\rm Li}$ & err & $A$(Li) & err   \\  
        & (K)   &  \multicolumn{2}{c}{(m\AA)}      &  \multicolumn{2}{c}{(m\AA)}  & \multicolumn{2}{c}{(dex)}   \\  
\hline
\noalign{\smallskip}
  &   &  \multicolumn{4}{c}{Measured values} \\
\noalign{\smallskip}
\hline
\noalign{\smallskip}
1    &  6070 & 101 & 15  & 107 &  6 & \dots & \dots  \\ 
2    &  4130 &  85 & 17  &  25 &  6 & \dots & \dots  \\ 
\noalign{\smallskip}
\hline
\noalign{\smallskip}
  &   &  \multicolumn{4}{c}{Corrected values} \\
\noalign{\smallskip}
\hline
\noalign{\smallskip}
1    &  6070 &  109 &  17 & 116 &  7 & 2.95 & 0.07  \\ 
2    &  4130 & 1209 & 247 & 361 & 81 & 1.97 & 0.35  \\ 
\hline
\end{tabular}
\end{center}
\label{Tab:Halpha_Lithium}
\end{table}

The spectral subtraction was also used to measure the excess emission in the core of the H$\alpha$ line ($W_{\rm H\alpha}^{em}$), which is an indicator of chromospheric activity.
As is apparent in Fig.~\ref{fig:subtraction_compo2}, the H$\alpha$ core of the
primary component is filled in by emission, while a redshifted H$\alpha$ emission bump is clearly visible at the wavelength corresponding to the secondary component. The net H$\alpha$ emission cores of the two components are  displayed by the hatched areas in the residual spectrum (blue line in the bottom panel of Fig.~\ref{fig:subtraction_compo2}). They were integrated to provide the values of $W_{\rm H\alpha}^{em}$, which are reported in Table~\ref{Tab:Halpha_Lithium}. We corrected these values for the flux contribution of the two stars to the observed spectrum, as we did for $W_{\rm Li}$.

\subsection{Out-of-eclipse variations}
\label{Subsec:modulation}

\begin{figure}
\hspace{-0.5cm}
\includegraphics[width=9.3cm]{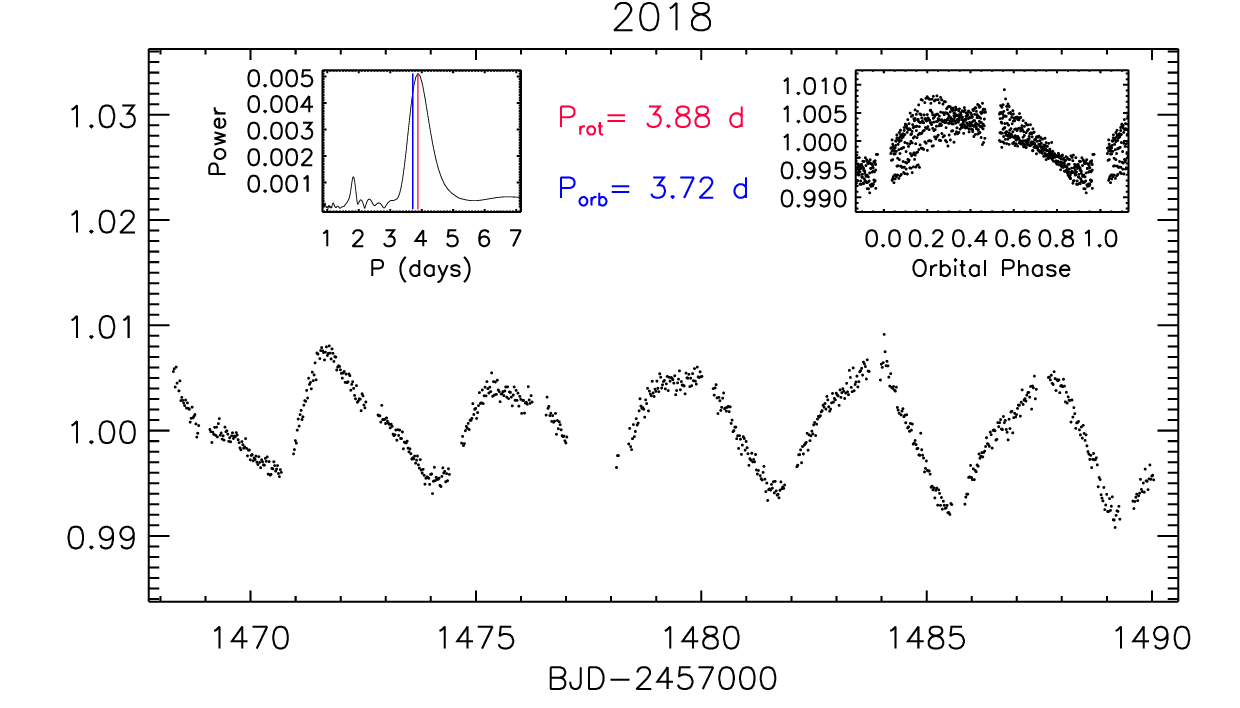}     
\vspace{0.cm}
\caption{{\it TESS} SAP light curve of \tic\ in 2018 (sector 6) where the eclipses have been removed. The inset in the upper left corner shows the cleaned periodogram; the rotational and orbital periods are marked with vertical red and blue lines, respectively. The inset in the upper right corner
displays the data phased with the orbital period.}
\label{fig:modul2018}
\end{figure}

\begin{figure}
\hspace{-0.5cm}
\includegraphics[width=9.3cm]{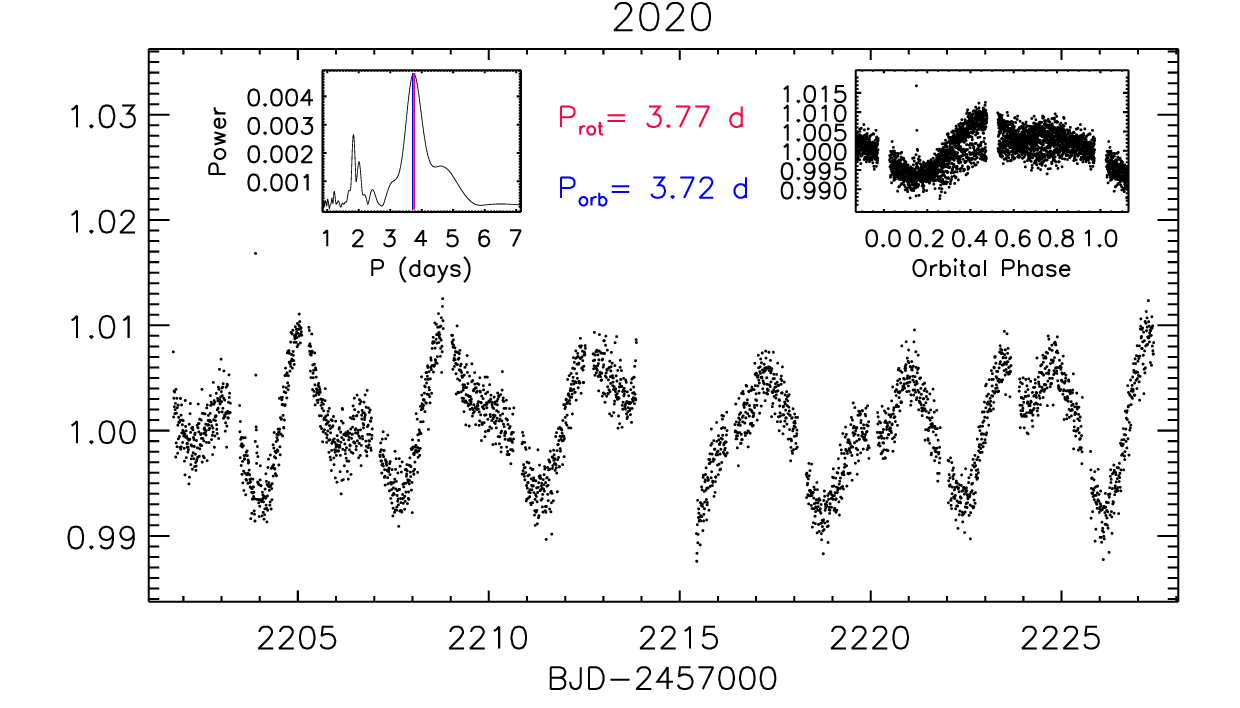}     
\vspace{0.cm}
\caption{Same as Fig.\,\ref{fig:modul2018} but for 2020 (sector 33) \tess\ SAP data.}
\label{fig:modul2020}
\end{figure}

The out-of-eclipse variations visible in the SAP fluxes (Fig.\,\ref{fig:TESS}) are reminiscent of rotational modulation produced by starspots in one or both of the system components. Indeed, proximity effects (ellipticity and reflection) are negligible for this very detached system 
(amplitude $\sim$\,3 mmag), as indicated by the synthetic light curve produced by \jktebop.
We applied a periodogram analysis \citep{Scargle1982} and the CLEAN deconvolution algorithm \citep{Roberts1987} to the {\it TESS} SAP light curves  
where the eclipses were removed.
The out-of-eclipse variations and the results of the period search are depicted in Fig.\,\ref{fig:modul2018} and Fig. \ref{fig:modul2020} for sectors 6 and 33, respectively.
The cleaned power spectrum displays a peak at a period very close to the orbital one in both \tess\ epochs (3.88$\pm$0.20 and 3.77$\pm$0.15 days, in 2018 and 2020, respectively).
This could be related to the rotation of the primary component, which is the component contributing the largest flux (more than 90\,\%) in the \tess\ bandpass. However, given the very low peak-to-peak amplitude of the observed modulation (15--20\,mmag), we cannot exclude a relevant contribution from the secondary component, which displays an active chromosphere in the H$\alpha$ line (see Fig.\,\ref{fig:subtraction_compo2}).
This suggests that at least one component has attained spin-orbit synchronization. 

The timescales for synchronization, 
calculated for the primary and secondary components according to \citet{Zahn1989}, are $\tau_{\rm sync}\sim 0.3$\,Myr and 8\,Myr, respectively. 
At NGC2232's age of 28 Myr, both components should already be synchronized or close to the spin-orbit synchronization, as found from the out-of-eclipse variations.

\section{Discussion}
\label{Sec:discussion}

The center-of-mass velocity of the \tic\ system ($\gamma$\,=\,26.3$\pm0.6$\,\kms, Table~\ref{Tab:Param}) is consistent with the cluster mean velocity RV=25.38$\pm$0.18\,\kms\ derived by \citet{Jackson2022} for 697 members of NGC\,2232 observed  as a part of the \gaia-ESO Survey (GES); it also agrees with the value of 25.35$\pm$0.85\,\kms\ based on \gaia\ DR2 data \citep{Soubiran2018}. 
This reinforces the membership of \tic\ to NGC\,2232.
We note that the GES observations in the field of NGC\,2232 did not include \tic.

From the corrected values of $W_{\rm Li}$ (Table~\ref{Tab:Halpha_Lithium}), we derived a lithium abundance, $A$(Li), of 2.95$\pm$0.07 and 1.97$\pm$0.35 for the primary and 
secondary component, respectively, by using the curves of growth of \citet{Lind2009}. The different lithium abundance for the two components follows the general 
trend of lithium depletion as a function of \teff\ displayed by clusters of similar ages \citep[e.g.,][]{Sestito2005}.
To estimate the age of the system from the lithium content in the photospheres of its components, we used the \eagles\ code \citep[][]{Jeffries2023}. This code fits  
Li-depletion isochrones to the values of \teff\ and $W_{\rm Li}$  of a coeval star group, such as the members of a cluster or the components of a binary system, 
as in the present case. 
We found $age=25^{+9} _{-19}$\,Myr (see Fig.~\ref{fig:EAGLES}), which is in very close agreement to the age of 28\,Myr estimated with \eagles\ by \citet{Jeffries2023} for 
the members of NGC~2232 observed as a part of the GES.

\begin{figure}[ht]
\hspace{-0.3cm}
\includegraphics[width=9.cm]{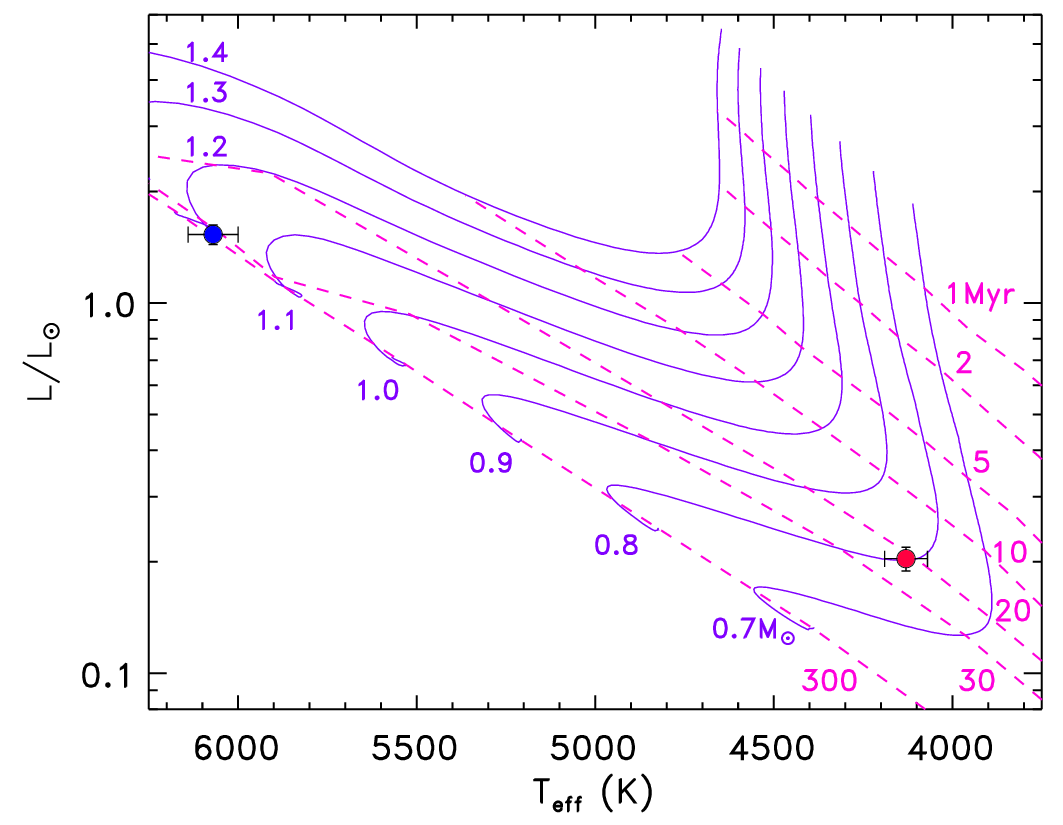}      
\caption{Position of the primary (blue dot) and secondary (red dot) component of TIC~43152097 in the HR diagram. Isochrones and evolutionary tracks from \citet{Baraffe15} 
are overlaid as dashed and solid lines; the labels give their age and mass.
}
\label{fig:HR}
\end{figure}

The position of the two components of TIC~43152097 in the Hertzsprung-Russell (HR) diagram is shown in Fig.~\ref{fig:HR}, along with the PMS evolutionary tracks and isochrones 
from \citet{Baraffe15}.
We note that the cooler component lies close to the isochrone at 20\,Myr and very close to the evolutionary track for 0.8\,\Msun, suggesting that it is slightly overluminous 
and a little bit hotter than a standard model would predict for a 0.76\,\Msun\ star with an age of $\simeq$30\,Myr. 
The primary component appears to have already reached the MS, since for this mass, all PMS isochrones with $age\geq$\,30\,Myr overlap the ZAMS ($age\simeq 300$\,Myr). 
It is very likely too hot to produce a strong magnetic activity and display significant radius inflation. Furthermore, its position in the HR diagram rules out an age $\leq20$\,Myr, 
at which the object should have  a noticeably higher luminosity and a lower temperature, according to the tracks and isochrones from \citet{Baraffe15}.

\begin{figure}[ht]
\hspace{-0.3cm}
\includegraphics[width=9.cm]{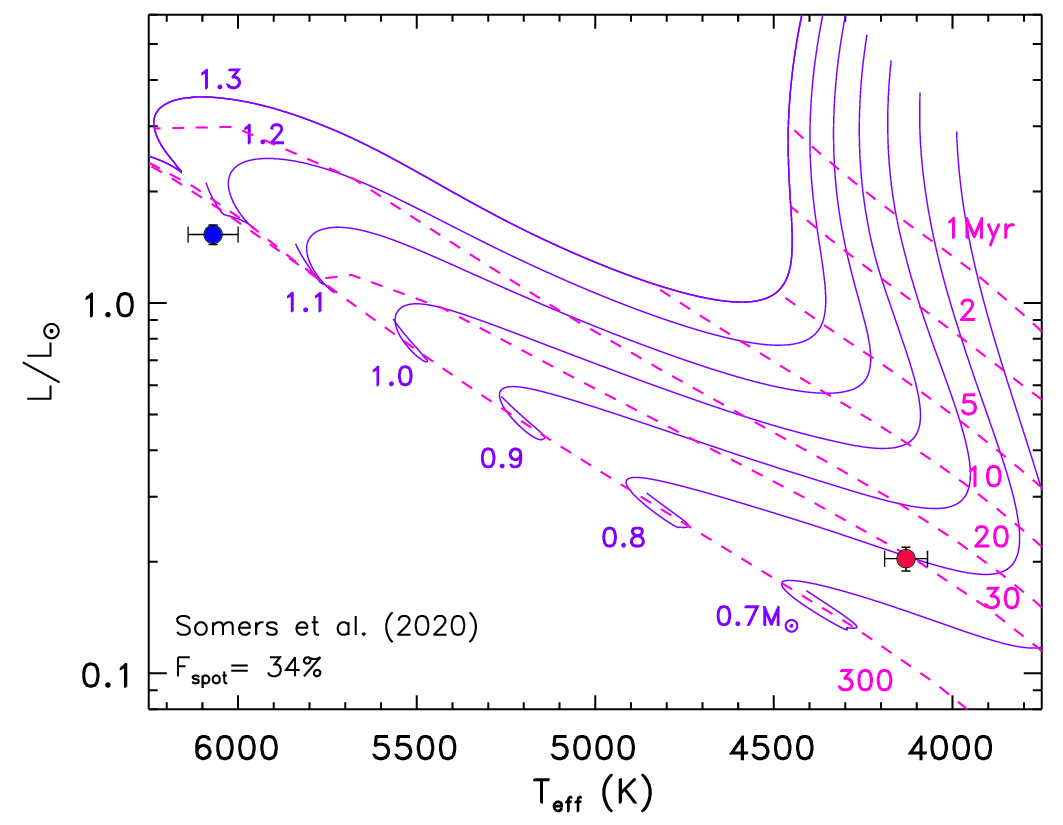}     
\caption{Position of the components of TIC~43152097 in the HR diagram. Isochrones and evolutionary tracks from \citet{Somers2020} for a spot covering fraction of 34\,\% are overlaid. Symbols and line styles are as in Fig.~\ref{fig:HR}.
}
\label{fig:HR_Somers}
\end{figure}

A better agreement between the position of the secondary component and the isochrone at 30\,Myr is found considering the effects of radius inflation (Fig.~\ref{fig:HR_Somers}). 
To this purpose, we used the SPOTS models \citep{Somers2020} that incorporate both the inhibition of convective energy transport caused by sub-photospheric magnetic fields and the 
effects of cool ($T_{\rm spot}/T_{\rm phot}=0.8$) starspots. 

\begin{figure}[ht]
\hspace{-0.3cm}
\includegraphics[width=9.cm]{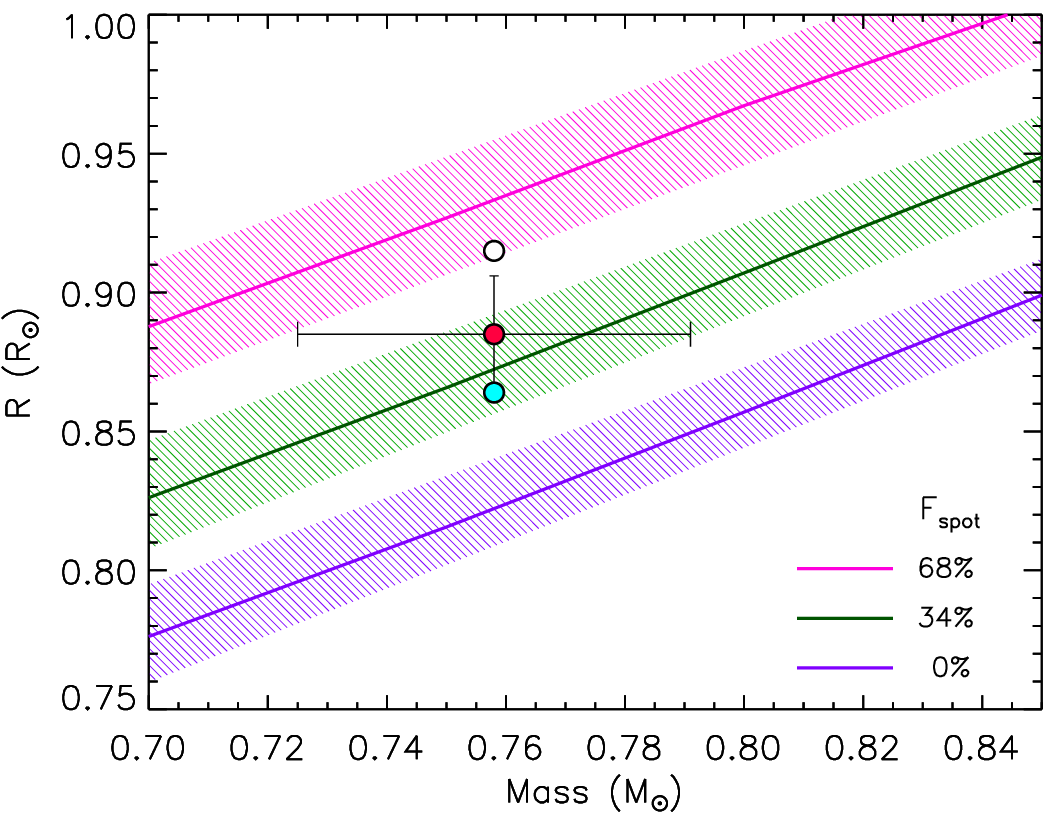}     
\caption{Mass--radius diagram for the secondary component of TIC~43152097 (red dot). SPOT isochrones \citep{Somers2020} at 28\,Myr ($\log(age)=7.45$) for a spot covering factor of 0\,\%, 34\,\%, and 68\,\% are overlaid with different colors, as indicated in the legend.  The hatched areas delimit, per each value of $F_{\rm spot}$, an age uncertainty of $\Delta\log(age)=0.05$. As a comparison, the value derived with \jktebop\ assuming a third light $L_3=0$ ($R_2=0.915$\,\Rsun) is also shown by the white dot. A larger spot coverage is necessary to reproduce the observations in this case. The value of $R_2=0.865$\,\Rsun, derived with \jktebop\ with the best-fit value of $L_3=-0.094$, is also overlaid with a cyan dot.
}
\label{fig:Mass_Radius}
\end{figure}

A more effective way to highlight the radius inflation for the secondary component is to compare $M_2$ and $R_2$ - quantities directly obtained from the light-curve and RV curve 
solutions - with the SPOT isochrones. 
To this purpose, we plotted the mass-radius diagram, which is displayed in Fig.\ref{fig:Mass_Radius}. In this diagram, the SPOTS isochrones at 28\,Myr ($\log(age)=7.45$) for 
three spot-covering factors (0\%, 34\% , and 68\,\%) are overplotted on the position of the secondary component.
For each spot coverage, we considered an age uncertainty $\Delta\log(age)=0.05$ (hatched strips in Fig.~\ref{fig:Mass_Radius}), which is a more conservative value compared to 
that of $\pm 0.02$ reported by \citet{Jeffries2023}. Figure~\ref{fig:Mass_Radius} clearly displays the discrepancy between the observed radius and the one predicted by models 
that neglect the effects of magnetic activity\footnote{The SPOT tracks and isochrones with a spot-filling factor $F_{\rm spot}= 0$\% are nearly identical to the tracks from 
\citet{Baraffe15}, as also shown by \citet{Frasca2021}.}. A radius inflation of approximately 8\,\% can be deduced. It is notable that a spot coverage of $\sim 40$\,\% is 
sufficient to explain the discrepancy, while a 68\,\% spot fraction would give rise to a too-high radius inflation.
If we adopt instead the slightly larger radius derived with \jktebop,\ assuming a third light $L_3=0$, the radius inflation, with respect to the isochrone without spots, would be approximately 11\,\%, which is compatible with a larger spot coverage 
($\sim 60\,\%)$. If we adopt the \jktebop\ solution with the best-fit value of $L_3$, we find a smaller value of radius inflation of about 7\,\%, compatible with a spot coverage of about 30\,\%.
This result indicates that radius inflation on the order of of 7--11\,\% is already occurring in late K-type stars at ages considerably earlier than the ZAMS, when these objects are still in the PMS evolutionary phase. 
The direct measure of the stellar radius for the components of PMS eclipsing binaries strongly supports radius inflation as the most likely cause of disagreement between the ages inferred from CMDs and the lithium depletion pattern in the PMS cluster Gamma Vel \citep{Jeffries2017}.
This, along with the problem of spectroscopic  \teff\ determination in heavily spotted stars, has a strong impact on the determination of ages and masses of PMS stars from their position in the HR diagram \citep[see, e.g.,][]{Gully-Santiago2017,Gangi2022}.

\section{Summary}
\label{sect:Conclusions}

We have reported the discovery and follow-up study of the eclipsing binary system \tic, which is composed of low-mass stars in a well-detached configuration and belongs to the young open cluster NGC\,2232. 

High-resolution spectra purposely collected with HARPS-N allowed us to measure the RV of the two components and determine their basic properties (atmospheric parameters, lithium abundance, and chromospheric emission).  
The RV of the center of mass of the system and the age derived from the photospheric lithium abundance fully agree with the cluster values, which confirms the membership of \tic\ to NGC\,2232.

We used the very precise \tess\ photometry to study the out-of-eclipse modulation and to determine the orbital period from the eclipse timing. The low-amplitude (15--20 mmag) modulation observed out of the eclipses shows a regular behavior with a period close to the orbital one and can be ascribed to cool starspots in one or both components, which are in a synchronous (or quasi-synchronous) rotation. 

The combined analysis of light and RV curves allowed us to derive orbital elements as well as fundamental stellar parameters of the two stars, particularly their masses and radii. The primary component of \tic\ is a late F-type dwarf (\teff\,=\,6070\,K, from the spectral analysis with \COMPO), while the lower-mass secondary is a late K-type PMS star (\teff\,=\,4130\,K). The precise measurements of \Rstar\ ($\simeq$2\,\%) and \Mstar\ ($\simeq$4\,\%)
indicate radius inflation for the secondary component, which turns out to be 7--11\,\% larger than predicted by standard evolutionary models. 
The SPOT evolutionary models, which incorporate both inhibition of convective energy transport caused by sub-photospheric magnetic fields and the effects of cool starspots covering a substantial fraction of the stellar surface (30--60\,\%), allow for reproduction of the stellar parameters of the secondary component at the age of NGC2232,  28 Myr. 

\begin{acknowledgements}
We are grateful to the referee, Dr. John Southworth, for his very useful comments and suggestions.
This research made use of SIMBAD and VIZIER databases, operated at the CDS, Strasbourg, France.\\
This paper includes data collected by the {\it TESS} mission which are publicly available from the Mikulski Archive for Space Telescopes (MAST). 
Funding for the {\it TESS} mission is provided by the NASA's Science Mission Directorate.\\
This work has made use of data from the European Space Agency (ESA)
mission \gaia\ ({\tt https://www.cosmos.esa.int/gaia}), processed by
the {\it Gaia} Data Processing and Analysis Consortium (DPAC,
{\tt https://www.cosmos.esa.int/web/gaia/dpac/consortium}). Funding
for the DPAC has been provided by national institutions, in particular the institutions participating in the {\it Gaia} Multilateral Agreement.\\
This research used the facilities of the Italian Center for Astronomical Archive (IA2) operated by INAF at the Astronomical Observatory of Trieste.
This work has been supported by the PRIN-INAF 2019 STRADE (Spectroscopically TRAcing the Disk dispersal Evolution), the Large-Grant INAF YODA (YSOs Outflow, Disks and Accretion), and by the Mini-Grant INAF 2022 ``High resolution spectroscopy of Open Clusters". 
X.F. thanks the support of the National Natural Science Foundation of China (NSFC) No. 12203100 and the China Manned Space Project with NO. CMS-CSST-2021-A08. 

\end{acknowledgements}

\bibliographystyle{aa}
\bibliography{bibliography}

\begin{appendix}

\section{Additional figures}
\label{sec:appendix}

\begin{figure}
\begin{center}
\hspace{-.7cm}
\includegraphics[width=9.3cm]{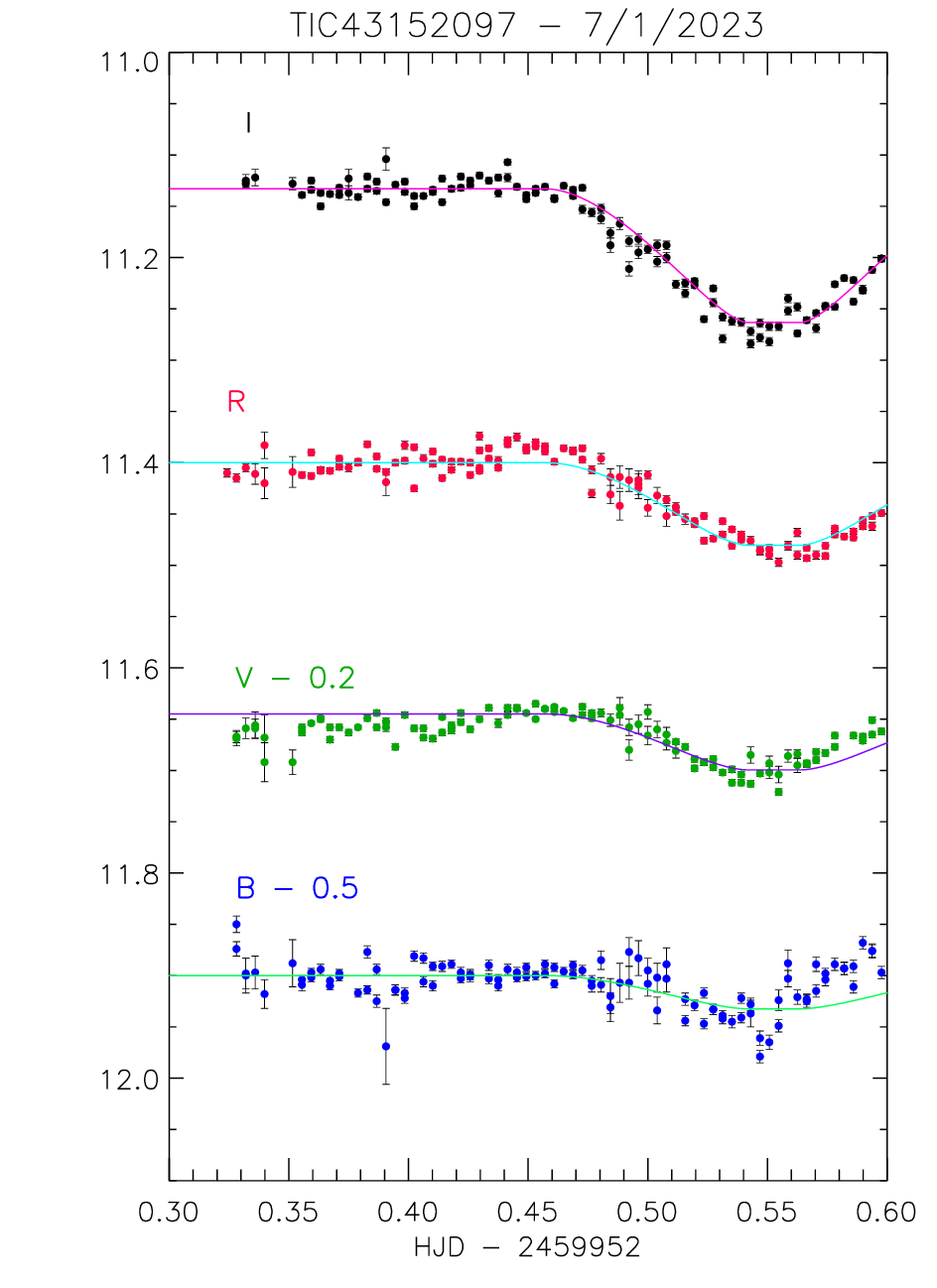}    
\caption{Multiband ground-based light curve of TIC~43152097 during the secondary eclipse of 7\ January 2023. The model light curves with the same 
elements as those listed in Table~\ref{Tab:LC_Param} are overplotted with solid lines.
}
\label{fig:ground_LC}
\end{center}
\end{figure}

\begin{figure}
\begin{center}
\hspace{-.5cm}
\includegraphics[width=9.cm]{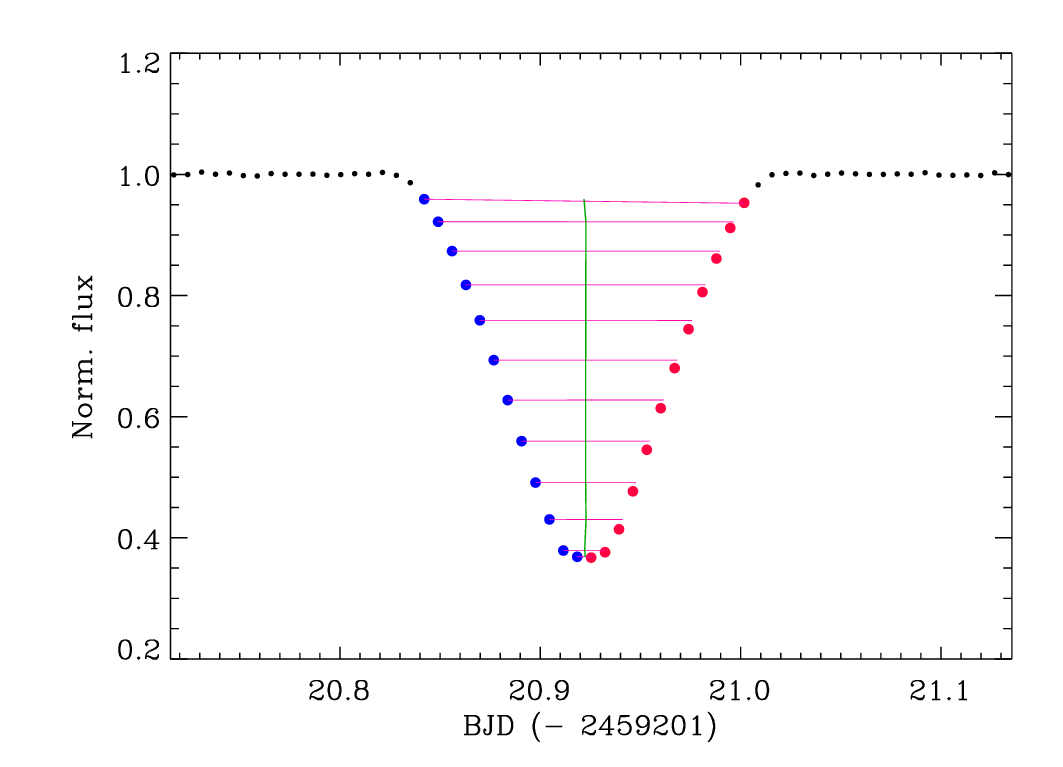}     
\vspace{0.cm}
\caption{Example of the determination of the time of a light-curve minimum with the bisector method for the primary eclipse observed by \tess\ in sector 33 on 7\ January 2021. 
The horizontal magenta lines connect points at the same flux level in the descending (blue dots) and rising (red dots) branch of the eclipse, interpolating between 
the two. The vertical green line represents the eclipse bisector.
}
\label{Fig:bisector}
\end{center}
\end{figure}

\begin{figure*}[t]
\includegraphics[width=6.cm]{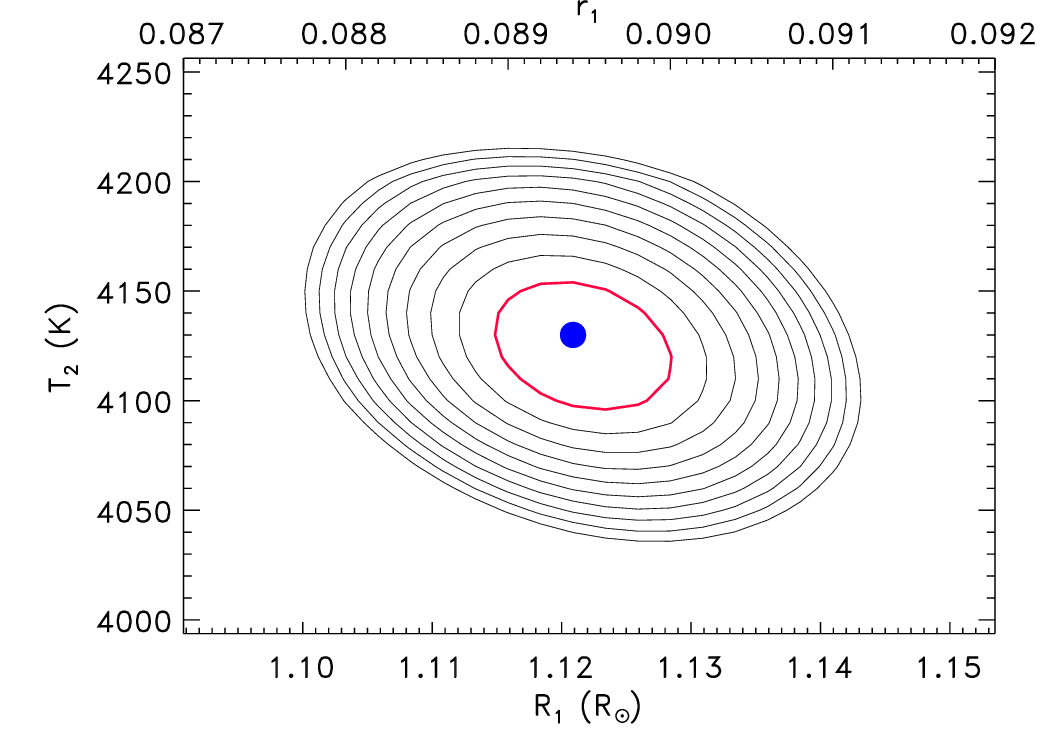}    
\includegraphics[width=6.cm]{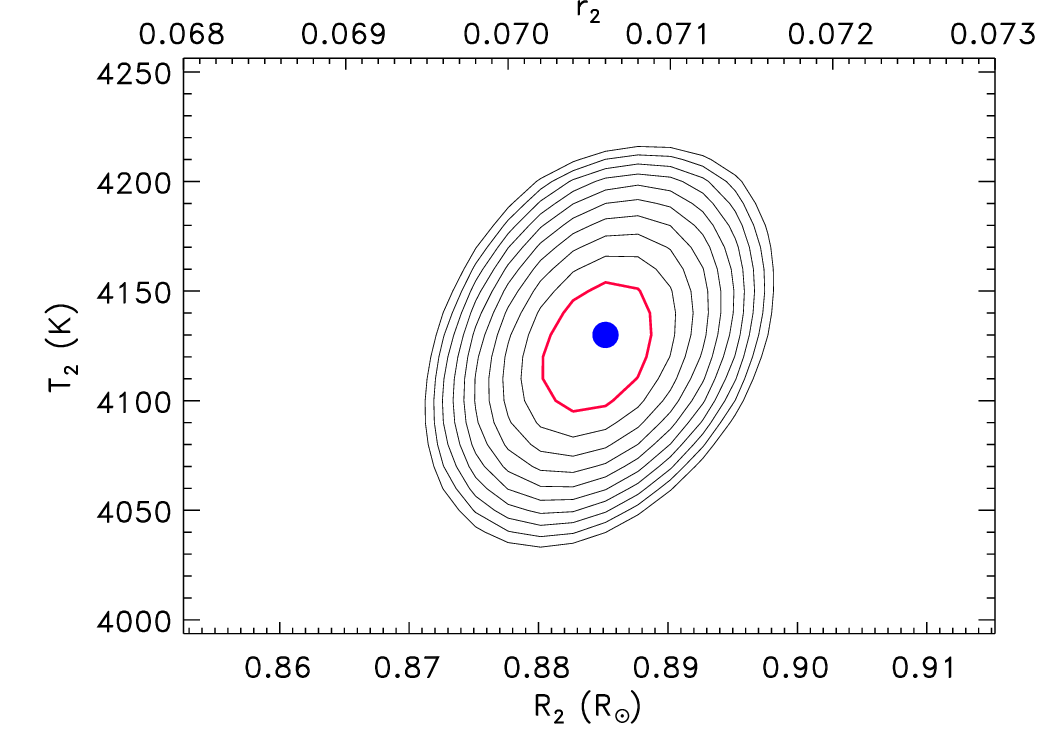}    
\includegraphics[width=6.cm]{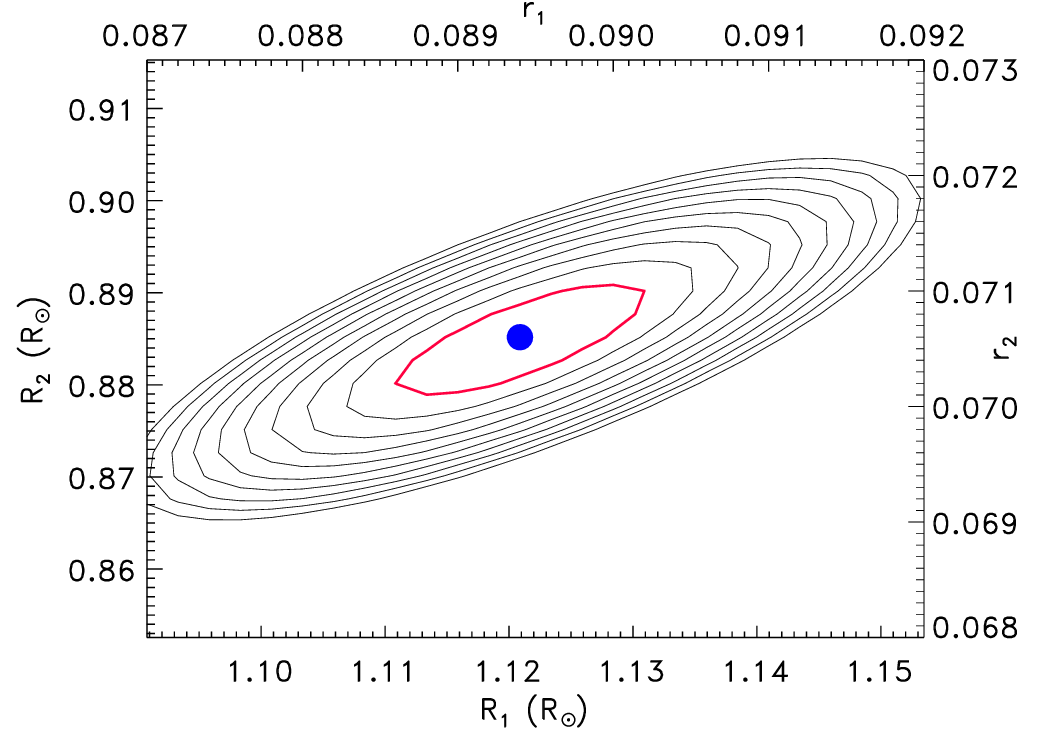}    
\includegraphics[width=6.cm]{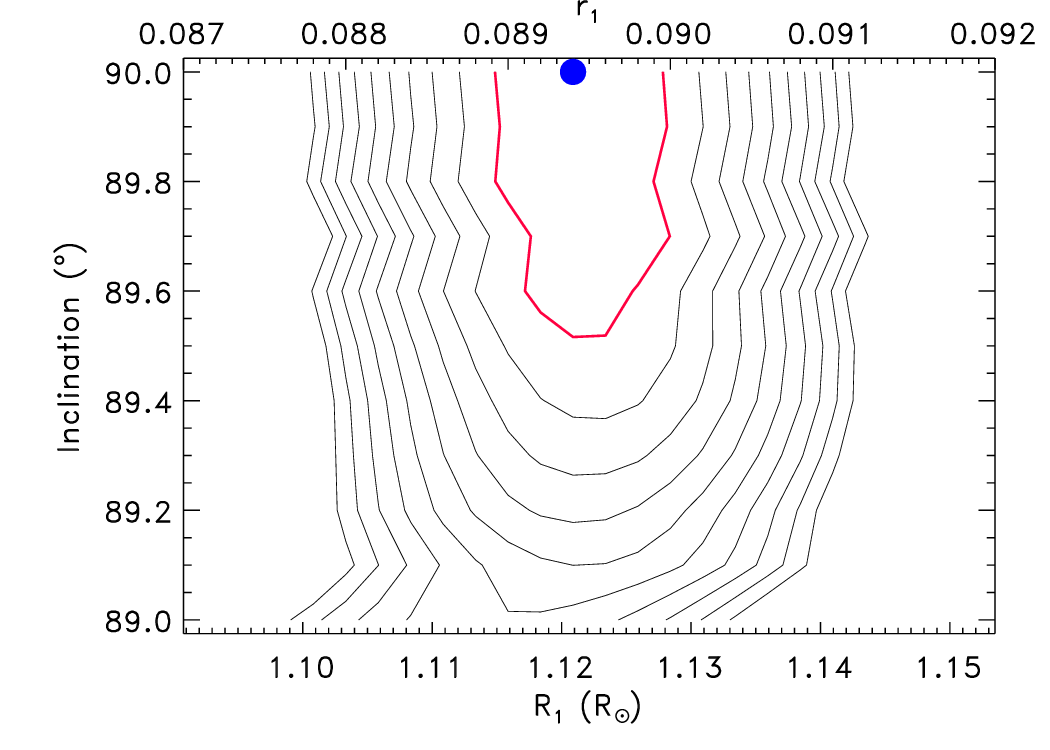}    
\includegraphics[width=6.cm]{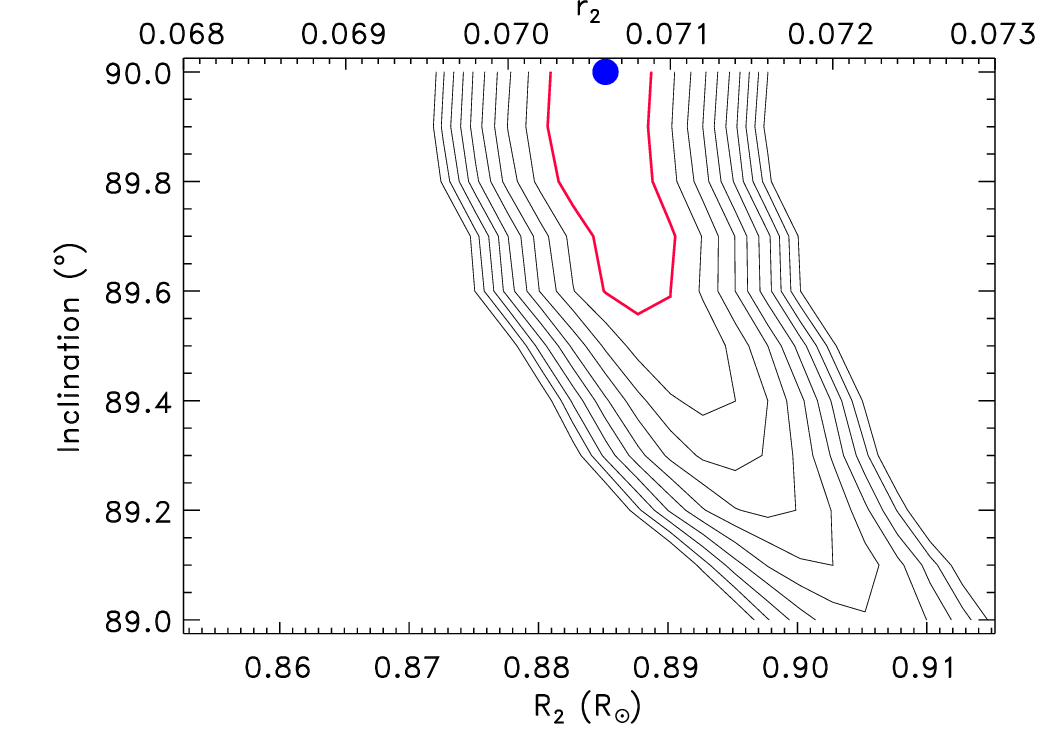}    
\includegraphics[width=6.cm]{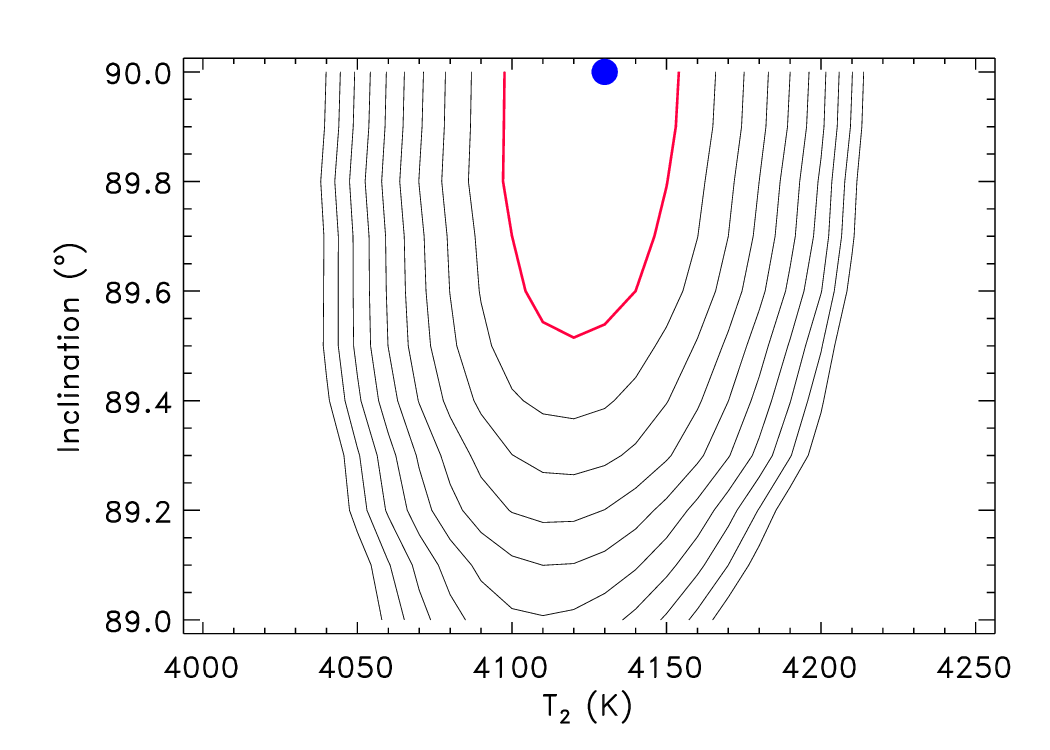}    
\caption{ $\chi^2$ contour maps for different pairs of parameters. In each panel, the blue dot marks the best-fit values,  while the 1$\sigma$ 
confidence level is denoted by the red contour. 
}
\label{fig:chi2}
\end{figure*}

\begin{figure}
\begin{center}
\hspace{-.7cm}
\includegraphics[width=9.3cm]{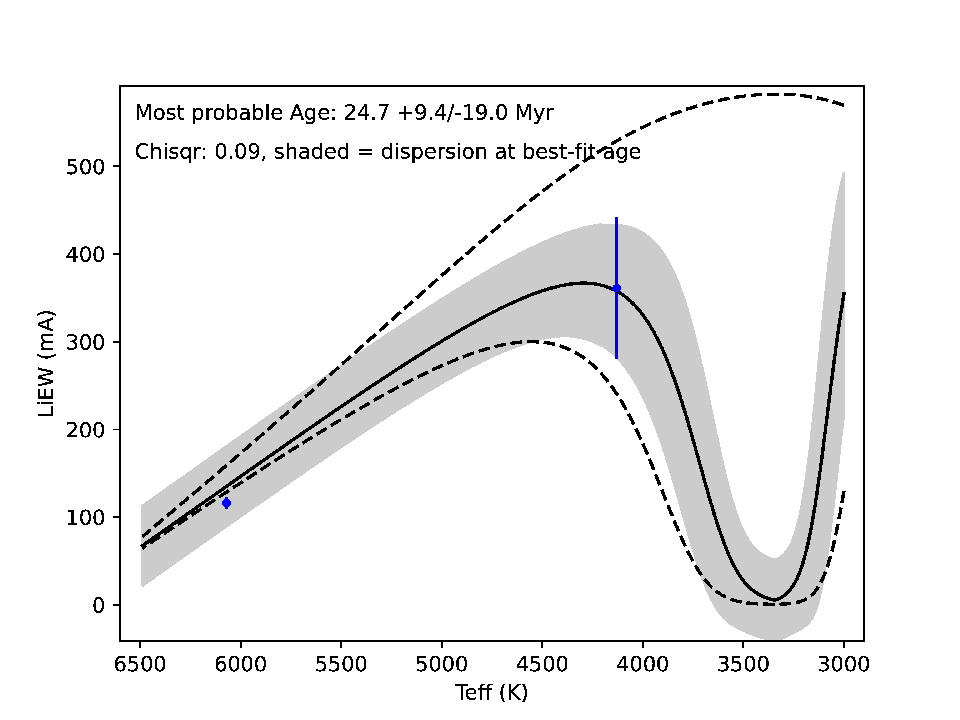}    
\caption{Fit to the lithium depletion pattern of the two components of \tic\ made with the \eagles\ code \citep{Jeffries2023}.
}
\label{fig:EAGLES}
\end{center}
\end{figure}

\end{appendix}

\end{document}